\documentclass{aastex}
\usepackage{emulateapj5}
\usepackage{apjfonts}
\journalinfo{Accepted for publication in The  Astrophysical Journal, 2003}
\slugcomment{Received 2002 October 31; accepted 2003 April 19}

\shorttitle{Super-Eddington Accretion Flow and Growth Timescale of BHs}
\shortauthors{KAWAGUCHI}

\def\brfrac#1#2{\left(\dfrac{#1}{#2}\right)}
\def\dfrac#1#2{{\displaystyle\frac{\mathstrut #1}{#2}}}

\def\ea{{\rm et~al.\ }}

\def\ltsima{$\; \buildrel < \over \sim \;$}
\def\lsim{\lower.5ex\hbox{\ltsima}}
\def\gtsima{$\; \buildrel > \over \sim \;$}
\def\gsim{\lower.5ex\hbox{\gtsima}}
\def\Mdot{{\dot{M}}}
\def\mdot{{\dot{m}}}
\def\Mbh{{M_{\rm BH}}}
\def\Msun{{M_\odot}}

\def\Ledd{{L_{\rm Edd}}}

\def\rs{{r_{\rm Sch}}}

\def\Qrad{{Q_{\rm rad}^-}}
\def\Qvis{{Q^+_{\rm vis}}}

\def\Hb{{{\rm H} \beta}}
\def\Fe2{{{\rm Fe} \sc II}}
\def\Tcol{{T_{\rm in}}}
\def\Tin{{T_{\rm in}}}

\begin{document}
\title{COMPTONIZATION \ IN \ SUPER-EDDINGTON \ ACCRETION \ FLOW \ AND 
\ GROWTH \ TIMESCALE \ OF \ SUPERMASSIVE \ BLACK \ HOLES}
\author{Toshihiro Kawaguchi}
\affil
{LUTH, Observatoire de Paris, Section de Meudon,
5 Place J.\ Janssen, 92195 Meudon, France}
\affil
{Department of Physics and Astronomy, The University of Oklahoma, 
440 West Brooks St., Norman, OK 73019, USA}
\affil
{Department of Astronomy, Graduate School of Science, 
Kyoto University, Sakyo-ku, Kyoto 606-8502, Japan}
\affil
{Postdoctoral Fellow of the Japan Society for the Promotion of
Science}

\authoremail{toshihiro.kawaguchi@obspm.fr}

\begin{abstract}
Super-Eddington accretion onto black holes (BHs) may 
occur at Ultra-Luminous
compact X-ray sources in nearby galaxies, Galactic microquasars
and narrow-line Seyfert 1 galaxies (NLS1s).
Effects of electron scattering (opacity and Comptonization) 
and the relativistic correction (gravitational redshift and
transverse Doppler effect)
on the emergent spectra from super-Eddington accretion flows 
onto non-rotating BHs
are examined for $10^{1.5}$ and $10^{6.5} \Msun$
BH masses ($\Mbh$). 
With $\mdot \, 
[\equiv \Mdot / (\Ledd / c^2)$, 
where $\Mdot$ is the accretion rate]$\, \geq 100$, the 
spectral hardening factor via electron scattering
is $\lsim 2.3 - 6.5$.
Due to the $\mdot$-sensitive hardening factor, 
the color temperature of the innermost radiation 
is not proportional to $L^{0.25}$, 
differing from the simplest standard accretion disk.
The model is applied to 
optical--soft X-ray emission from NLS1s.
We pick up one NLS1, namely PG~1448+273 with an inferred $\Mbh$ of 
$10^{6.4} \Msun$, among the highest $\mdot$ candidates.
The broadband spectral distribution is
successfully reproduced by the model 
with an extremely high $\mdot$ ($=$1000) and the viscosity parameter 
$\alpha$ of 0.01. 
This implies that this object, as well as some other highest $\mdot$
systems, is really young: 
the inferred age, $\Mbh / \Mdot$, is about $10^6$ years.
We also briefly discuss the distribution of $\mdot$ 
for transient and highly variable NLS1s, finding that
those are located at $3 \lsim \mdot \lsim 300$.
Such a moderately high accretion rate is indicative of
thermal instability.
Furthermore, $\mdot$ for a possible type-2 counterpart of NLS1s, NGC~1068,
is found to be similar to $\mdot$ for NLS1s.
\end{abstract}

\keywords{accretion, accretion disks --- black hole physics ---
galaxies: active --- galaxies: nuclei --- X-rays: galaxies ---
 X-rays: stars}

\section{INTRODUCTION}

Super-Eddington accretion phenomena onto black holes (BHs) 
may be important for formation and evolution 
of supermassive BHs.
It may have been much more common in the 
young universe, when the average black hole mass ($\Mbh$) 
would have been smaller than now.
Such phenomena are possible sources of UV--Extreme UV (EUV) photons that
may have governed the thermal and star formation history
of the universe.
Moreover, those extreme accretion rate phases provide us a laboratory
in examining our knowledge on accretion phenomena, which 
have been studied intensively 
under sub-Eddington regimes (e.g., Shakura \& Sunyaev 1973).

Some Ultra-Luminous compact X-ray sources (ULXs) and Galactic 
microquasars
show large bolometric luminosity ($\gsim$ the Eddington luminosity 
$\Ledd$ of 1.4 $\Msun$) and 
high color temperature ($\sim$ 1~keV), 
often referred as $T_{\rm in}$
(e.g., Roberts \& Warwick 2000; 
Makishima \ea 2000; Colbert \& Ptak 2002).
We note that recent results of ULXs obtained with {\it XMM-Newton} seem to
show lower $T_{\rm in}$ ($\sim$0.15~keV; Miller \ea 2003).
Some possible physical explanations for their nature, 
including rotating BHs, high accretion rate 
($\Mdot > \Ledd / c^2$, where $\Mdot$ is
the gas accretion rate), intermediate BH mass ($\sim$100$\Msun$) and
beamed radiation (see, however, Misra \& Sriram 2003) have been discussed.

There is a subclass of Active Galactic Nuclei (AGNs) 
that are candidates for super-Eddington accretion,
namely narrow-line Seyfert 1 galaxies (NLS1s) 
and their luminous counterpart, narrow-line QSOs.
They are characterized as follows (see Pogge 2000).
(i) They have narrower Balmer lines of hydrogen 
(e.g., FWHM of $\Hb$ $\leq$ 2000 km s$^{-1}$), relative to
usual broad-line Seyfert 1 galaxies (BLS1s) and QSOs having
FWHM of $\Hb$ $\gsim$ 5000 km s$^{-1}$ (e.g., Osterbrock \& Pogge 1985).
(ii) They often emit strong optical 
\Fe2 multiplets (e.g., Halpern \& Oke 1987).
(iii) Their optical--soft X-ray, big blue bump is hotter than BLS1s/QSOs:
they have bluer optical spectra (Grupe et al. 1998) 
and often show steep and luminous soft X-ray
excess (Pounds et al. 1996; Otani et al. 1996; Boller et al. 1996; 
Wang \ea 1996; Laor et al. 1997; Leighly 1999b).
(iv) Rapid soft/hard X-ray variability is another characteristic 
of NLS1s (Otani et al. 1996; Boller et al. 1997; 
Leighly 1999a; Hayashida 2000).
The soft X-ray spectral variability of NLS1s is opposite to
that of BLS1s: 
spectra get harder when they become brighter (Cheng, Wei \& Zhao 2002).

When the luminosity from the accretion disk/flow $L$ becomes close to $\Ledd$
[$
\sim 1.3 \times 10^{45} (\Mbh / 10^7 \Msun) \, 
{\rm erg} \ {\rm s}^{-1}$],
it has been shown that the 
advective energy transport in the flow dominates over
radiative cooling (Abramowicz et al. 1988;
section 10.3 of Kato, Fukue \& Mineshige 1998).
Such a disk is called an the optically-thick ADAF 
(advection-dominated accretion flow) or slim disk, 
since it is moderately thick geometrically.  
Abramowicz \ea showed that $L$ saturates at around a few
times $\Ledd$ when $\Mdot$ exceeds $\Ledd / c^2$, 
and that the flow shines even inside the marginally
stable orbit [$3 \, \rs$ for non-rotating BHs; where $\rs$ is the 
Schwarzschild radius and  
$\rs \equiv 2 G \Mbh  / c^2 \simeq 3 \times 10^{12} (\Mbh / 10^7 \Msun)$
cm = $10^{-3} \, (\Mbh / 10^7 \Msun)$ lt-days.].
The physical reason why the advection turns on at super--Eddington 
regime will be the fact that photons can not escape from the flow
before photons/gas are swallowed by the central BH (the photon trapping
effect; Begelman \& Meier 1982; we discuss it in \S 6.2).
Slim disk models for ULXs have been constructed to explain 
High $\Tcol$ (Watarai \ea 2000) 
and mysterious behaviors in $\Tcol$ v.s.\ $L_{\rm X}$ diagrams 
(Watarai, Mizuno \& Mineshige 2001).
Similarly, these models are successful in discussing strong soft 
X-ray emission (large $\Tcol$ relative to other AGNs) and
time variability of NLS1s (Mineshige \ea 2000, with $\Mbh$ and $\Mdot$ 
being free parameters).
Currently, the most promising picture of NLS1s and narrow-line QSOs is
that they contain
relatively less massive black holes (with $\Mbh \sim 10^{6-8}M_\odot$) and
higher $L/\Ledd$ (with $L \sim \Ledd$).

As to the emergent spectra from 
super-Eddington accretion, Szuszkiewicz, Malkan \&  Abramowicz (1996)
and Wang et al. (1999) considered the opacity due to electron scattering,
but without Comptonization nor the relativistic correction.
Mineshige \ea (2000) and Watarai \ea (2000) 
simply apply the local blackbody approximation,
whereas Watarai \ea (2001) take the spectral hardening factor
of 1.7 from Shimura \& Takahara (1995) in order to 
mimic the effect of Comptonization.
We note that this factor is derived in the sub-Eddington 
regime, and thus the extent of Comptonization at super-Eddington
phase is not evaluated by it (see \S 3.2).
Self--irradiation and --occultation are discussed by
Fukue (2000) with self--similar solutions of the slim disk model 
(e.g., Begelman \& Meier 1982).
Most recently, Wang \& Netzer (2003) have studied the effect of 
Comptonization with self-similar solutions, 
in order to explain X-ray emission from NLS1s.

A big problem remains when we compare the slim disk model with the 
observed optical/UV--soft X-ray spectra of NLS1s.
Namely the predicted soft X-ray is too step compared with
observations (Szuszkiewicz \ea 1996; Wang \ea 1999).

We examine the effects of electron scattering in opacity, 
of Comptonization and of the relativistic correction 
upon spectra from transonic slim accretion flows
onto non-rotating BHs, 
in order to figure out to what extent those effects change the 
emergent spectrum and whether the effects are enough to fit 
the observed spectral energy distributions.
Now that $\Mbh$ estimations for AGNs are accumulating,
we are able to compare the models and observational
data with less free parameters than previous studies.
We have obtained a promising fit to the broad-band energy distribution
for the first time.
Methods and assumptions of the numerical calculations are given in \S 2.
We then discuss the physical quantities and emergent spectra 
of the slim disk in \S 3.
We turn to the comparison of the model with
observed, broad-band spectra of NLS1s in the next section.
The growth timescale of supermassive BHs is discussed in \S 5.
In \S 6, several discussions are presented.
The final section is devoted to a summary.
Throughout the present study we use the normalization 
$\Mdot_{\rm Edd} (\equiv L_{\rm Edd}/c^2$)
for $\Mdot$:
\begin{eqnarray}
 \Mdot_{\rm Edd}
   &\simeq& 1.3\times 10^{18} \, (\Mbh / 10   \Msun)~ {\rm g~s}^{-1}   \cr
   &\simeq& 1.3\times 10^{24} \, (\Mbh / 10^7 \Msun)~ {\rm g~s}^{-1}   \cr
   &\simeq& 2.0\times 10^{-2} \, (\Mbh / 10^7 \Msun)~ \Msun {\rm ~yr}^{-1}.
\end{eqnarray}
Hereafter, $\mdot$ refers to $\Mdot / \Mdot_{\rm Edd}$.

Throughout this study, 
the luminosity is calculated from observed flux assuming 
isotropic radiation, zero cosmological constant, 
deceleration parameter $q_0 = 0.5$, and Hubble constant $H_0 = 75$ 
km/s/Mpc.

\section{Numerical Procedures}

\subsection{The Structure of the Accretion Disk}

In order to discuss the shape of the emergent spectra from the 
innermost regions, 
it is essential to solve the full, derivative equations,
rather than the self-similar solutions or other approximate solutions
(based on a priori inner boundary conditions or 
no regularity conditions).
The transonic nature of the accretion flow should also be
carefully treated.

We solve the steady-state, transonic accretion disk structure.
The numerical methods are basically the same as those adopted in 
Matsumoto et al. (1984).
The code has been extended for time-dependent simulations 
(Matsumoto, Kato \& Honma 1989; Honma \ea 1991a,b)
and it has been made use of in subsequent papers;
Takeuchi (2000), Watarai \ea (2000, 2001) and Mineshige \ea (2000).
Here, we summarize the numerical procedures, and briefly describe
the basic equations in the following paragraphs. 
We employ the pseudo-Newtonian potential, 
$\psi = -GM/(r-\rs)$ (Paczy\'{n}ski \& Witta 1980),
and cylindrical coordinates, ($r,\varphi,z$).

In this study, we use vertically integrated (height-averaged) derivative 
equations.
The scale-height of the disk $H$ is comparable to radius $r$ 
in the case of extremely large accretion rates 
$\Mdot \gsim 100 \, \Ledd /c^2$.
One may wonder if such a height-averaged approach would not be appropriate.
The same problem arises in an optically thin 
ADAF (Narayan \& Yi 1995; 
see section 10.2 of Kato et al. 1998 and references therein).
Narayan \& Yi (1995) found that 
the 2-dimensional solutions of the exact non-height-integrated 
equations agree quite well with those of the simplified 
height-integrated equations 
when the `height-integration' is done along a constant spherical
radius, rather than along $z$ at constant cylindrical 
radius (see Narayan 1997). 
The height-integrated equations therefore are fairly accurate
descriptions of quasi-spherical, advective flows.

We construct the vertically integrated
equations, using the integrated variables, such as 
surface density, $\Sigma \equiv \int\rho dz = 2 \bar{\rho} H$, 
and integrated total (gas plus radiation) pressure,
$\Pi \equiv \int pdz = 2 \bar{p} H$.
Throughout this study (except for the absorption coefficient $\bar{\kappa}$), 
each quantity with a bar on top means the vertically 
averaged value, evaluated from the value at the mid plane 
(value with the suffix ``mid''):
$\bar{\rho} = (16/35) \, \rho_{\rm mid}$,
$\bar{p} = (128/315) \, p_{\rm mid}$ and 
$\bar{T} = \frac{2}{3} \, T_{\rm mid}$, respectively 
(H\={o}shi 1977; see Matsumoto \ea 1984).
The scale-height of the flow
$H$ is determined by hydrostatic balance
between the vertical gravity and pressure:
i.e. $H = \Omega_{\rm K}/c_s$, where
$c_s$ is the sonic speed.
Equations for the conservations of the 
mass, momentum, and angular-momentum are
\begin{equation}
\label{mass}
   -2\pi r\Sigma v_r={\dot M}= {\rm const.},    
\end{equation}
\begin{equation}
\label{r-mom}
   v_r{dv_r\over dr}+{1\over \Sigma}{d\Pi\over dr}
   ={\ell^2-\ell_{\rm K}^2\over r^3}
   -{\Pi\over \Sigma}{d{\rm ln}\Omega_{\rm K}\over dr},   
\end{equation}
and
\begin{equation}
\label{ang-mom}
    {\dot M}(\ell-\ell_{\rm in})=-2\pi r^2T_{r\varphi},  
\end{equation}
respectively.
Here,
$\Omega (=v_\varphi/r)$, $\Omega_{\rm K}
  [=(GM/r)^{1/2}/(r-\rs)$],
$\ell$ ($=rv_\varphi$),
$\ell_{\rm K}$ ($\equiv r^2\Omega_{\rm K}$) and $\ell_{\rm in}$
are respectively, the angular frequency of the gas flow,
the Keplerian angular frequency
in the pseudo-Newtonian potential,
the specific angular momentum,
 the Keplerian angular momentum and
the specific angular momentum finally 
swallowed by the BH.
As to the viscous stress tensor we adopt the usual prescription
$T_{r\varphi} \equiv -\alpha \Pi$ with the viscosity
parameter $\alpha$.
Unless Otherwise noted, we apply $\alpha$ of 0.1.
We neglect self-gravity of the disk, for simplicity.
We will see that this is self-consistent with the computed density
profile.

The energy budget (per unit surface area) at each radius is symbolically 
written as
\begin{equation}
\label{energy}
       Q_{\rm adv}^-=Q_{\rm vis}^+-Q_{\rm rad}^-,  
\end{equation}
where $Q_{\rm adv}^- [\propto \Sigma v_rT(ds/dr)$] 
is the advective cooling with $s$ being specific entropy,
and the other two terms on the right-hand side represent
viscous heating and radiative cooling, respectively.
Assuming that the effective optical depth $\tau_{\rm eff}$
is larger than unity,
we apply the usual, diffusion approximation for $Q_{\rm rad}^-$ 
(cf. \S 3.1);
\begin{equation}
\label{eq:qrad}
       Q_{\rm rad}^- = \frac{8 a c T_{\rm mid}^4}{3 \bar{\kappa} 
\rho_{\rm mid} H}
\approx \frac{H u_{\rm rad}}{{\rm t}_{\rm diff}},
\end{equation}
where the total opacity coefficient is $\bar{\kappa} = 
\bar{\kappa_{\rm abs}} + \bar{\kappa_{\rm es}}$.
Here, $u_{\rm rad}$ and t$_{\rm diff}$ mean 
the radiation energy density per unit volume ($a \bar{T}^4$)
and the diffusion time scale of photons (see \S 3.1), respectively.
This equation describes the photon leakage.
Neutrino leakage in proto-neutron stars is discussed 
in \S 6.2.
If $\tau_{\rm eff}$ gets smaller than unity, 
$u_{\rm rad}$ will be lower than $a \bar{T}^4$.
The current cooling rate (eq.~\ref{eq:qrad})
will then overestimate the true cooling rate (Beloborodov 1998).
An improvement of the formula, e.g., multiplying the cooling rate
by $\left(1-\exp^{-\tau_{\rm eff}}\right)$, would be required in the future.

The outer boundary conditions are imposed at 
$r = 2.0 \times 10^4 \rs$, where
each physical quantity is taken from the formula of
the standard disk (Shakura \& Sunyaev 1973).
The choice of the outer radius does not influence on the
inner solutions (e.g., Abramowicz \ea 1988).
The basic equations are integrated by the semi-implicit method
from the outer boundary 
to the inner one taken at $r = 2.5 \rs$ for 
$\mdot \leq 10$ and 1.01 $\rs$ for $\mdot \geq 100$.
The solutions satisfy the regularity condition at the
transonic radius, at about 2.7$\rs$,
and we adopt the free boundary conditions at the inner edge.

The system is taken to be a disk consisting of 
thermal plasma around a Schwarzschild black hole of mass $\Mbh$.
We also assume 1-Temperature gas (i.e., temperatures of 
protons and of electrons are the same).
The Rosseland mean opacity is adopted for the absorption opacity 
$\bar{\kappa_{\rm abs}}$; 
$\bar{\kappa_{\rm abs}} = \bar{\kappa_{\rm abs, 0}} \,
\bar{\rho} \, \bar{T}^{-3.5}$.
In this study, every optical depth is measured between 
the surface and the mid-plane of the flow (not from one surface to the other).
Generally, the optical depth of absorption $\tau_{\rm abs}$ 
($H \bar{\rho} \bar{\kappa_{\rm abs}}$)
is much less than that of electron scattering $\tau_{\rm es}$
($H \bar{\rho} \bar{\kappa_{\rm es}}$, with the electron scattering 
opacity $\bar{\kappa_{\rm es}}$ of 0.4 cm$^2$/g)
in the region of interest ($r \lsim 10^{3-4} \rs$; 
cf.\ \S 3.2 in Kato \ea 1998).
The flow is optically thick; $\tau_{\rm abs} + \tau_{\rm es} \gg 1$.
The absorption opacity is an important factor when we are considering
whether the effective optical depth of the flow 
[$\tau_{\rm eff} \equiv \sqrt{3 \tau_{\rm abs} 
(\tau_{\rm abs}+\tau_{\rm es})}$; Rybicki \& Lightman 1979] 
exceeds unity or not. 
To estimate the absorption coefficient 
$\bar{\kappa_{\rm abs, 0}}$ (mainly bound-free 
transitions for AGN disks with 
solar abundances; see e.g., Laor \& Netzer 1989), 
we simply take 30 times the free-free absorption coefficient 
($\bar{\kappa_{\rm ff, 0}} = 6.4~10^{22}$ cm$^2$/g) for 
the super massive BHs, 
as is done in Czerny \& Elvis (1987).
For much smaller $\Mbh$ (32 $\Msun$), we adopt 
$\bar{\kappa_{\rm ff, 0}}$ as $\bar{\kappa_{\rm abs, 0}}$.
The timescale of Comptonization and that of 
Coulomb collisions are much shorter
than the accretion timescale (Beloborodov 1998).

We neglect the gas evaporation from the disk 
(e.g., Meyer \& Meyer-Hofmeister 1994), since it becomes weak
when the gas accretion rate approaches the Eddington limit (Liu \ea 2002).
The coronae above and beyond the disk are not included in this 
study.
It will be necessary to consider them when applying the models 
to hard X-ray spectra (Wang \& Netzer 2003).

In total, the input parameters required for the calculations are
$\Mbh$, $\Mdot$ and $\alpha$. 

\subsection{Spectral Calculations}

We follow Czerny \& Elvis (1987) and Wandel \& Petrosian (1988), 
in dealing with the effect of opacity of electron scattering (i.e. 
the modified blackbody) and the effect of  the energy exchange via Compton 
scattering (Comptonization).
More precise treatments, 
including such as the effects of radiative transfer 
(see Shimura \& Manmoto 2003), will be 
required for more detailed comparison of models and observations.
Given the $T_{\rm eff}$, the local spectrum $I_\nu (r)$ can be expressed by 
$I_\nu (r) = B_\nu(T) f_\nu(T,r)$, where $B_\nu(T)$ is the Planck function.
Here, $f_\nu(T,r)$ is 
\begin{eqnarray}
\label{f_nu_mbb}
 f_\nu(T,r) = \frac{ 2 \, [1-\exp^{- \tau_{\rm eff}(T,r)}] }
  { 1 + \sqrt{ [\tau_{\rm abs, \nu}(T,r)+\tau_{\rm es}(r)] /
                     \tau_{\rm abs, \nu}(T,r) }} ,
\end{eqnarray}
where $\tau_{\rm abs, \nu}(T,r)$ is the optical thickness for absorption 
with a given temperature $T$ at frequency $\nu$, and the 
effective optical depth $\tau_{\rm eff, \nu}(T,r)$ 
is computed as 
{\footnotesize
$\sqrt{3 \tau_{\rm abs, \nu}(T,r) 
[ \tau_{\rm abs, \nu}(T,r) + \tau_{\rm es}(r) ]}$}.
The parenthesis including the exponential term takes into account 
the finite, effective optical depth of the flow.
For low $\tau_{\rm eff, \nu}(T,r)$ ($\ll 1$), it describes the thermal
bremsstrahlung from an optically thin medium.
If the absorption opacity dominates over the opacity of electron 
scattering (and if the flow is effectively, optically thick), 
$f_\nu(T,r)$ is unity.
The color temperature $T$ is determined so that the surface emissivity 
does not change: $\int^{\infty}_{0} d\nu I_\nu (r) 
\equiv a c T_{\rm eff}^4 / 4 \pi$.

A simple evaluation of the spectral hardening factor due to
the modified blackbody is (Madej 1974):
\begin{eqnarray}
\label{eq_madej}
\log_{10}\brfrac{T_{\rm col}}{T_{\rm eff}} \simeq - \frac{1}{8} 
\log_{10}\brfrac{\tau_{\rm abs}}{\tau_{\rm abs}+\tau_{\rm es}}.
\end{eqnarray}

Compton scattering is important when the Compton parameter 
{\footnotesize $y = (4 k \bar{T} / m_e c^2)$
Max($\tau_{\rm es}, \tau_{\rm es}^2)$} is greater than 1.
However, below the last thermalization surface, where the effective
optical depth measured from the surface equals unity, 
the energy exchange via Compton scattering will be statistically
canceled due to the efficient absorption.
Scattering above the surface is in question.
Then, the relevant opacity for electron scattering, $\tau_{\rm es}'$,
will be $\tau_{\rm es} / \tau_{\rm eff}$ for $\tau_{\rm eff} > 1$ and
$\tau_{\rm es}$ for $\tau_{\rm eff} < 1$:
\begin{eqnarray}
 \tau_{\rm es}' = \frac{\tau_{\rm es}}{{\rm Max}(\tau_{\rm eff},1)}.
\end{eqnarray}
Finally, the effective Compton-$y$ parameter $y_*$ is
\begin{eqnarray}
\label{eq_eff_Comp}
 y_* = \frac{4 k \bar{T}}{m_e c^2} 
       {\rm Max}(\tau_{\rm es}', \tau_{\rm es}'^2).
\end{eqnarray}
Comptonization will considerably modify the emergent spectrum if
$y_* > 1$.
The radial distribution of $y_*$ increases inwards steeply, reaching
a maximum at $\sim (3-5) \rs$, and it decreases inward 
(Wandel \& Petrosian 1988).
In the regime for which 
$\tau_{\rm es} \gg \tau_{\rm abs}$ and $\tau_{\rm eff} > 1$,
\begin{eqnarray}
\label{eq_y_mdot_dep}
y_* \propto \bar{T}^{4.5} \bar{\rho}^{\, -1} 
 \brfrac{\bar{\kappa_{\rm abs, 0}}}{\bar{\kappa_{\rm ff, 0}}}^{-1}, 
\end{eqnarray}
yielding a strong temperature dependency.

\begin{figure*}[tb]
\figurenum{1}
\centerline{\includegraphics[angle=0,width=12cm]{f1.epsi}}
\figcaption{\footnotesize
Opacity of electron scattering ($\tau_{\rm es}$, solid lines),
that of absorption ($\tau_{\rm abs}$, dotted lines) and 
the effective opacity ($\tau_{\rm eff}$, dot-dashed lines).
Here, $\Mbh$ and $\alpha$ are chosen to be $10^{6.5} \Msun$ and 0.1,
respectively.
It is shown that $\tau_{\rm es}$ is 
larger than $\tau_{\rm abs}$ by several orders 
at $r \lsim 10 \, \rs$ for all the $\mdot$ (1-1000).
The left panel is drawn for $\mdot = 10$ and 1000, while
the right one for $\mdot = 1$ and 100.
Open circles indicate the position where $p_{\rm rad} = p_{\rm gas}$.
Inner parts are $p_{\rm rad}$-dominated regions.
With $\mdot = 1000$, entire region presented here are $p_{\rm rad}$-dominated.
\label{fig:tau_m6_5}}
\end{figure*}

Photons generated at large optical depth undergo many scatterings
and change their energy as a result of Comptonization.
The effect on emergent spectra can be estimated by assuming
that a fraction $f_{\rm th, \nu}$ of the photons at frequency $\nu$
 are shifted and cumulated into a Wien peak of an average energy $3 k T$.
For a given frequency $\nu$, the fraction of these thermalized photons
to all generated photons $f_{\rm th, \nu}$ is given as 
\begin{eqnarray*}
f_{\rm th, \nu} = \hspace{7cm}
\end{eqnarray*}
\vspace{-3mm}
\begin{eqnarray}
\label{f_nu_Comp}
\exp 
\left\{ - \frac{\ln (k T / h \nu)}{\tau_{\rm es}^2 \ln 
 [1 + 4 k T / m_e c^2 + 16 (k T / m_e c^2)^2]} \right\}
\end{eqnarray}
(Svensson 1984), if the absorption is negligible.
For high-energy photons ($k T < h \nu$), $f_{\rm th, \nu}$
would be able to exceed unity.
We therefore put a constraint so that $f_{\rm th, \nu} \leq 1.0$.
Since the redistribution of low-energy photons into 
Wien peak is the problem, this artificial constraint on high-energy
photons will be a reasonable approximation (cf.\ Wandel \& Petrosian 1988).
The formula (eq.~\ref{f_nu_Comp}) can be adapted to 
a more general case if $\tau_{\rm es}$
in the equation is replaced by the total optical depth 
$\tau_\nu^{\rm total}$ (Czerny \& Elvis 1987):
\begin{eqnarray}
\tau_\nu^{\rm total} &=& \frac{\tau_{\rm es} + \tau_{\rm abs,\nu}}
                            {1+\tau_{\rm eff}} \cr
 &\approx& \tau_{\rm es} \qquad \qquad
    ({\rm for} \ \tau_{\rm eff} \ll 1, \ {\rm and} \ 
     \tau_{\rm es} \gg \tau_{\rm abs}) \cr 
 &\approx& \tau_{\rm es}^{1/2} \tau_{\rm abs}^{-1/2} \ \ \ \ \ \, 
    ({\rm for} \ \tau_{\rm eff} \gg 1, \ {\rm and} \ 
     \tau_{\rm es} \gg \tau_{\rm abs}).
\end{eqnarray}
The removed photons will result in a total additional 
energy output 
(we use $2.7kT$ instead of $3kT$, following Wandel \& Petrosian 1988)
\begin{eqnarray}
I_{\rm th, \nu} = 2.7 k T \int^{\infty}_{0} d\nu f_{\rm th, \nu} 
 \frac{I_\nu}{h \nu}.
\end{eqnarray}
The distribution of the shifted photons around
the Wien peak is approximated by a blackbody spectrum
with a normalization constant $C$ given by
\begin{eqnarray}
I_{\rm th, \nu} \approx C a c T^4 / 4 \pi.
\end{eqnarray}
As a whole, the deviation of the emitted spectrum from
a blackbody is finally described as 
\begin{eqnarray*}
 f_\nu(T,r) = \hspace{7cm}
\end{eqnarray*}
\vspace{-3mm}
\begin{eqnarray}
\label{eq_final_f_nu}
 \frac{ 2 \, [1-\exp^{- \tau_{\rm eff}(T,r)}] }
  { 1 + \sqrt{ [\tau_{\rm abs, \nu}(T,r)+\tau_{\rm es}(r)] /
                     \tau_{\rm abs, \nu}(T,r) } }
  [1 - f_{\rm th, \nu}] + C,
\end{eqnarray}
instead of $f_\nu(T,r)$ in eq.~(\ref{f_nu_mbb}).
Again, the temperature $T$ ($\geq T_{\rm eff}$) is adjusted to 
reproduce the original surface emissivity, 
so that the luminosity of the flow is conserved.
As the result, the emergent spectra become harder than 
spectra based on the local blackbody approximation [$B_\nu (T_{\rm eff})$;
Czerny \& Elvis 1987; Wandel \& Petrosian 1988].
A hot surface layer will arise whose temperature $T$ is much larger
than $T_{\rm eff}$, as seen in standard disks
with $\mdot > 1$ (Shimura \& Takahara 1993).

Next, we add the effects of the relativistic correction on 
the effective temperature $T_{\rm eff}$
and emergent spectra.
Since only the gravitational redshift and 
transverse Doppler shift (i.e.  no Doppler boosting) are considered, 
the computational results correspond to face-on views of the flow.
We assume that the rotational velocity $v_\varphi$ is almost the same as 
the Keplerian velocity $v_{\rm K}$.
The consistency will be shown later 
with the calculated velocity field (\S~3.1).
We employ the relativistic correction for the Keplerian disk 
(e.g., Kato \ea 1998, \S3.5.3);
\begin{eqnarray}
\label{eq:rc}
 (1+z_{r.c.}) = \left(1 - \frac{3 \rs}{2 R} \right)^{-1/2}.
\end{eqnarray}
We take $1+z_{r.c.}$ = [Max(10$^{-3}$, $1 - 3 \rs / 2 R$)]$^{-1/2}$,
for the moment.
The effective temperature $T_{\rm eff}$ 
becomes lower by $(1+z_{r.c.})^{-1}$ 
[i.e., emissivity per surface area decreases by $(1+z_{r.c.})^{-4}$],
compared with the cases without the relativistic correction 
(e.g., Watarai \ea 2000, 2001; Mineshige \ea 2000).

We have applied this expression to describe the shape of 
the radiation spectrum emitted at every radius of the disk.

\section{NUMERICAL RESULTS}

In this section, we show the basic physical quantities
in the accretion flow and emergent spectra for $\Mbh$ of $10^{6.5} \Msun$
and $\alpha = 0.1$.
For AGNs ($\Mbh \gsim 10^6 \Msun$), $\Mbh$ is now estimated by 
a couple of methods,
and both of the inner and outer radii are observable at
X-ray and optical/UV bands.
We chose $\alpha$ of 0.1 in order to compare the $T_{\rm eff}(r)$ profile
and the resultant spectra with those in Mineshige \ea (2000).
We also describe briefly the 
$\alpha$-dependency using the $\alpha = 0.01, \ 10^{-3}$ results,
as well as the $\Mbh$-dependency comparing 
with $10^{1.5} (=32) \, \Msun$ BH results.

\subsection{Physical Quantities of the Flow}

Figure \ref{fig:tau_m6_5} exhibits the opacity of 
electron scattering ($\tau_{\rm es}$),
that of absorption ($\tau_{\rm abs}$) and 
the effective opacity ($\tau_{\rm eff}$).
Here, $\Mbh$ and $\alpha$ are chosen as $10^{6.5} \Msun$ and 0.1,
respectively.
The figure clearly shows that $\tau_{\rm es}$ (solid lines) 
at $r \lsim 100 \rs$ is larger than $\tau_{\rm abs}$ (dotted lines) 
by several orders for all the $\mdot$ (1-1000).
The deviation is much more prominent in the super-Eddington
phases than the sub-Eddington phase ($\mdot \lsim 10$).
Therefore, the local blackbody approximation must be modified,
especially in the super-Eddington cases where $\tau_{\rm es}$ 
enormously dominates over $\tau_{\rm abs}$ (see eq.\,[\ref{eq_madej}]).
With $\mdot \sim 100$, $\tau_{\rm eff} \leq 1$ at $r \lsim 4 \rs$.
Thus, the diffusion approximation for the cooling rate (eq.\ \ref{eq:qrad})
will not be valid around this $\mdot$.
We will therefore search for objects with even higher $\mdot$
in \S 4.1.
The surface density $\Sigma$ is proportional to
$\tau_{\rm es}$: $\Sigma$ decreases with an increasing $\mdot$
as far as $\mdot \leq 100$.
As $\mdot$ increases from 100 to 1000, $\Sigma$ conversely increases
(since $v_{\rm r}$ increases only slightly) and the density 
at the inner region ($r \lsim 50 \rs$) is also enlarged:
both of $\tau_{\rm es}$ and $\tau_{\rm abs}$ get larger.
Consequently, the effective optical depth with $\mdot =$
1000 recovers to above unity.
Open circles exhibit the radii where the radiative 
pressure ($a \bar{T}^4 / 3$) equals to the gas pressure
($2 k_{\rm B} \bar{\rho} \bar{T} / m_{\rm H}$).
The former dominates over the latter within the inner region.
The $\mdot$-dependency of the radius is similar to
the standard disk ($\propto \mdot^{16/21}$).

We present the effective Compton-$y$ parameter $y_*$ 
(eq.~\ref{eq_eff_Comp})
for various accretion rates.
Each solid line in figure \ref{fig:y_m6_5} 
shows the distribution of $y_*$ as a 
function of radius $r$.
The horizontal dotted line means $y_*$ of unity, above which 
the spectral distortion due to Comptonization is crucial.
The large increase of $y_*$ along with the increase of $\mdot$ 
arises from a
decrease of $ \bar{\rho}$
and an increase of $\bar{T}$ from $\mdot = 10$ to 100--1000 
(see fig.~\ref{fig:hdtt_m6_5}).
Since $\bar{\rho} \propto \Mbh^{-1}$ and
$\bar{T} \propto \Mbh^{-1/4}$ (as will be shown later),
it is quite difficult to achieve a large 
$y_* (\propto \bar{T}^{4.5} \bar{\rho}^{\, -1}; 
{\rm eq.}~\ref{eq_y_mdot_dep})$ with sub-Eddington
accretion rates for all relevant $\Mbh$.
In other words, the huge $y_*$ ($\gg 1$) is obtainable 
only for super-Eddington cases.

\vspace{0.5cm}
\figurenum{2}
\centerline{\includegraphics[angle=0,width=8.5cm]{f2.epsi}}
\figcaption{\footnotesize
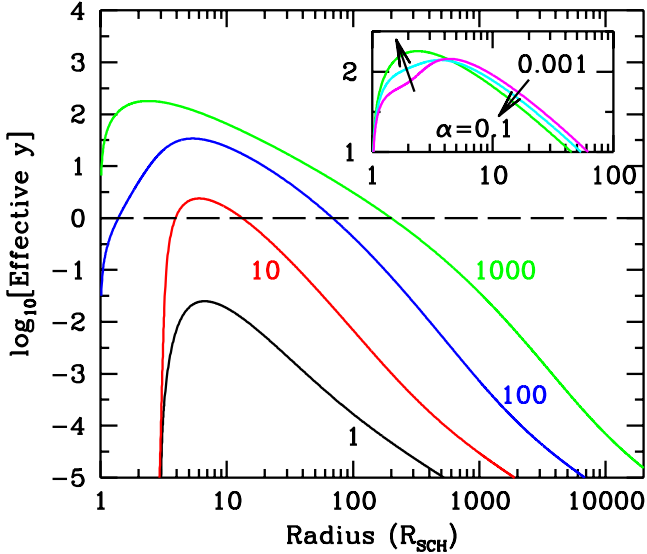
Effective Compton-$y$ parameter, $y_*$, for various $\mdot$.
A long-dashed line means that $y_* = 1$, above which Comptonization
is very important on the spectral calculations.
It is clear that $y_*$ with super-Eddington accretion rate is 
much larger than the cases with sub-Eddington accretion.
The inset shows the $\alpha$-dependency of $y_*$
for $\mdot = 1000$:
$\bar{\rho}$ increases with smaller $\alpha$, then
$y_*$ at the innermost region decreases.
\label{fig:y_m6_5}}
\vspace{0.5cm}

We assume Keplerian rotation for the transverse 
Doppler effects (\S 2.2).
Figure \ref{fig:vv_m6_5} shows the transverse velocity $v_\phi$ is 
almost equivalent to the Keplerian velocity $v_{\rm K}$ 
($v_\phi \gsim 0.8 \, v_{\rm K}$) up to the extremely large 
$\mdot$ (1000) cases.
Thus, the assumption is not very bad.
The radial velocity $v_{\rm r}$ increases with 
a larger $\mdot$ (lower panel).
This is the reason behind the ignition of the photon trapping in the
super-Eddington accretion.
In the extreme super-Eddington cases ($\mdot \gsim 1000$), 
$v_{\rm r}$ in the innermost region 
($r \lsim 10 \rs$) 
is about 10\% of the Keplerian one 
(i.e., $v_{\rm r} \simeq \alpha v_{\rm K}$).
Thus, the inner flow could be more time variable than lower $\mdot$ cases.
In the extreme limit of advection dominated flow 
($Q_{\rm adv}^- \gg Q_{\rm rad}^-$), the radial velocity $v_{\rm r}$
is expected to be $\sim 0.4 \alpha v_{\rm K}$ 
(e.g., Fukue 2000).
Filled dots indicate the location of the sonic point, where
$v_{\rm r}$ equals to the sound speed $c_{\rm s}$.
The sonic points are located around $3 \rs$ for all $\mdot$,
being consistent with the $\alpha=0.1$ cases in Abramowicz \ea (1988).

\vspace{0.5cm}
\figurenum{3}
\centerline{\includegraphics[angle=0,width=8.5cm]{f3.epsi}}
\figcaption{\footnotesize
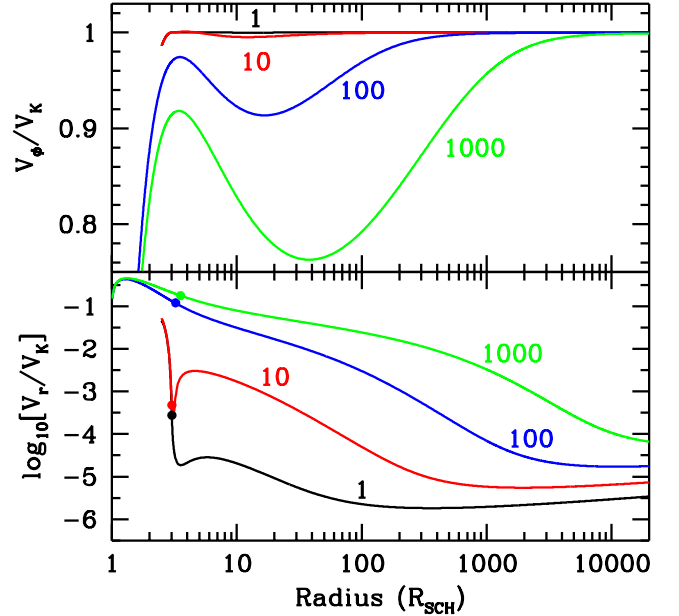
Azimuthal velocity (upper) and radial velocity (lower) 
in the unit of the Keplerian velocity $v_{\rm K}$ as 
functions of radius.
The rotation is almost Keplerian for all the $\mdot$ here,
within $\sim$20~\% deviation at most.
The radial velocity increases drastically with increasing
$\mdot$:
it reaches $\sim$10~\% of the Keplerian velocity for $\mdot = 1000$ 
at inner region ($r \lsim$ 10$\rs$).
Filled dots indicate the location of the sonic points 
($v_{\rm r} / c_{\rm s} = 1$) for
each parameter set, where the sonic speed $c_{\rm s}$ is calculated 
as $\sqrt{\bar{p} / \bar{\rho}}$.
\label{fig:vv_m6_5}}
\vspace{0.5cm}

\begin{figure*}[tb]
\figurenum{4}
\centerline{\includegraphics[angle=0,width=12cm]{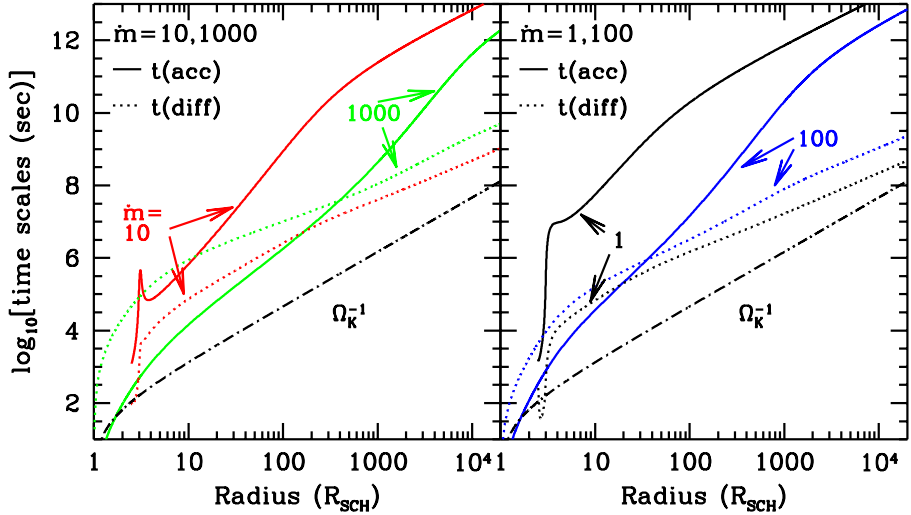}}
\figcaption{\footnotesize
Accretion timescale 
(t$_{\rm acc} = (r-\rs)/v_{\rm r}$; solid lines) and 
diffusion timescale of photons 
(t$_{\rm diff} = H \, (\tau_{\rm es} + \tau_{\rm abs})/c$; dotted)
for the slim accretion disk model.
The viscosity parameter $\alpha$ is 0.1 here.
The switch of the two time scale happens at $\mdot$ between 10 and 100.
The dot-dashed lines mean the vertical, hydrodynamical timescale 
[$H/c_{\rm s} = \Omega_{\rm K}^{-1} = 
255 \, (\Mbh / 10^7 \Msun) (r / 3 \rs)^{0.5} (r / \rs - 1)$\,sec
for the Pseudo-Newtonian potential].
\label{fig:t_scales_m6_5}}
\end{figure*}

Next, we compare the timescales of accretion 
[${\rm t}_{\rm acc}$ defined as $(r - \rs)/v_{\rm r}$] with that of 
diffusion of photons from mid-plane to the surface of the disk
[${\rm t}_{\rm diff} \equiv H (\tau_{\rm es} + \tau_{\rm abs})/ c$].
Figure \ref{fig:t_scales_m6_5} shows that 
${\rm t}_{\rm acc}$ is larger than ${\rm t}_{\rm diff}$ everywhere
for $\mdot \leq 10$, while it is not always true for 
$\mdot \geq 100$.
The switch of the two timescales happens between $\mdot$ of 
10 and 100.
Advective cooling (photon trapping) begins to 
take place from the inner region.
For example, the outer part ($r > 40 \, \rs$ for $\mdot =100$ and
$R > 400 \, \rs$ for $\mdot =1000$) does not
suffer from the effects of photon trapping.
Those radii can be analytically derived as $0.5 \, \mdot$ 
(Begelman \& Meier 1982).
It is also clear that the accretion timescale at a
fixed radius decreases drastically with increasing $\Mdot$.
In super-Eddington cases, the structure of the inner accretion 
flow can be highly time variable.
The timescale for achieving the vertical hydrostatic balance,
$H/c_s = \Omega_{\rm K}^{-1}$, 
is always much shorter than t$_{\rm acc}$.

The scale height of the flow $H$ in the unit of radius $r$ is
shown in figure \ref{fig:hdtt_m6_5} (top panel).
The aspect ratio $H/r$ increases with $\Mdot$, being
geometrically thick ($H/R \gsim 0.5$)
at $\mdot \gsim 100$.
We shall discuss this issue later when we are seeking ideal objects to 
test the model (\S 4.1).
Moreover, the shape of $H/R$ shows that the heating by the irradiation from
the innermost part of the flow onto the outer radii
($\gsim 10^4 \rs$) is negligible.
The radial profile
of the mean density $\bar{\rho}$ is also shown 
(solid lines in the second panel).
As Beloborodov (1998) noted, 
$\bar{\rho}$ at $r \lsim 50 \rs$ has a minimum at $\mdot \sim 100$.
Here, the dot-dashed line indicates a critical density, $\rho$(sg,z), above 
which self-gravity of the disk in the vertical (z) direction must
be considered: 
\begin{eqnarray*}
\rho({\rm sg,z}) 
=\Omega_{\rm K}^2 / (4 \pi {\rm G}) = 18 \, (\Mbh / 10^7 \Msun)^{-2} 
(r / 3 \rs)^{-1}
\end{eqnarray*}
\vspace{-3mm}
\begin{eqnarray}
\label{eq:rho_sg}
 \hspace*{4cm} (r / \rs - 1)^{-2} \, {\rm g/cm}^3.
\end{eqnarray}
As $\mdot$ increases, the relative importance of self-gravity 
[$\rho ({\rm sg,z}) / \bar{\rho}$] in the outer region is enhanced.
With $\mdot > 1000$, the outer part of the flow 
[$r \gsim 2 \times 10^4 \rs \simeq 0.006 (\Mbh / 10^{6.5} \Msun)$pc]
will form a self-gravitating disk.
Objects with super-Eddington accretion rate are therefore
good laboratories for studying the
self-gravity of accretion flows by optical/Near-IR observations.
Quantitative details will be shown in a future paper (Kawaguchi,
Pierens, \& Hur\'{e} 2003).
Another critical density for radial self gravity, $\rho$(sg,r) is
$16 \rho$(sg,z) (Goldreich \& Lynden-Bell 1965; see 
Hur\'{e} 1998 for a review).
The inertia of the heat (Beloborodov, Abramowicz \& Novikov 1997; 
dashed lines in the second panel) 
is always much smaller than the gas rest mass density $\bar{\rho}$
in the this study (disks around non-rotating BHs):
$\rho_{\rm rad} \equiv a \bar{T}^4 / c^2 
\simeq 10^{-11} (\bar{T} / 10^6 {\rm K})^4 \, {\rm g/cm^3} 
\ll \bar{\rho}$ (third panel).

Finally, the effective temperature profiles $T_{\rm eff} (r)$ 
before the relativistic correction (solid lines)
are compared with those with the correction (dotted lines; eq.~\ref{eq:rc}).
Although the flow with $\mdot > 100$ extends much closer to the central BHs
than lower $\mdot$ cases, 
those regions are not so luminous as thought in the previous papers 
(Watarai \ea 2000, 2001; Mineshige 2000).
If the system is not face-on, the discrepancy of the two profiles 
shrinks due to the Doppler boosting and the reduction of transverse
Doppler effect.
Viewing-angle dependent spectra are needed for investigation in the future.
The dashed line also shows $T_{\rm eff} (r)$, but it is derived 
without advection (i.e., $\Qvis = \Qrad$):
\begin{eqnarray*}
T_{\rm eff} (r)
 = 6.2 \ 10^5 \brfrac{\Mbh}{10^7 \Msun}^{-1/4}
   \brfrac{\Mdot}{\Ledd / c^2}^{1/4} 
\end{eqnarray*}
\vspace{-1mm}
\begin{eqnarray}
 \hspace*{2cm}    \brfrac{r}{\rs}^{-3/4}
       \left(1 - \sqrt{\frac{3 \rs}{r}} \right)^{1/4} {\rm K}.
\end{eqnarray}
%
\figurenum{5}
\centerline{\includegraphics[angle=0,width=8.2cm]{f5.epsi}}
\figcaption{\footnotesize
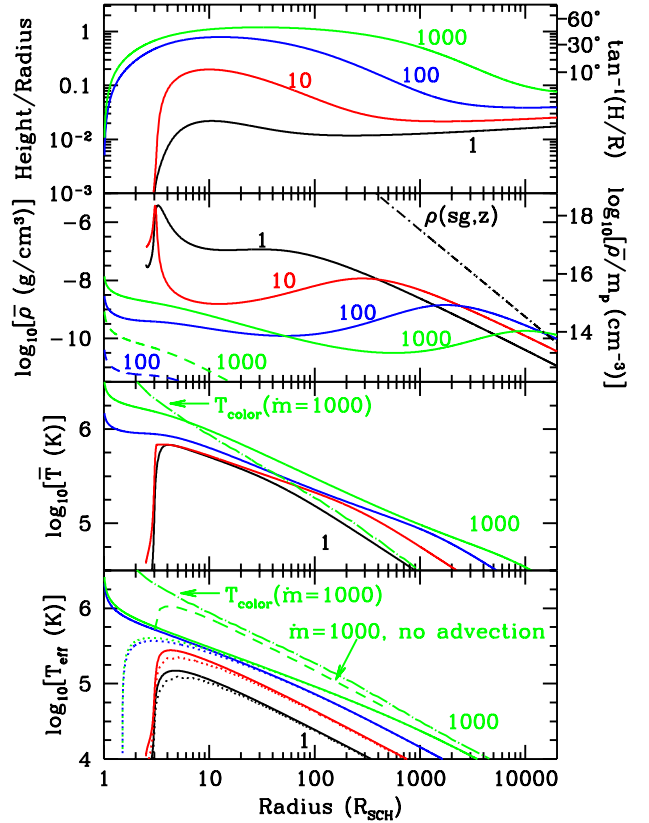
Examples of $\mdot$-dependency of the physical quantities.
The corresponding $\mdot$ are labeled.
First, the scale height of the flow (from the mid-plane to the surface) 
divided by the radius is shown (top).
The second panel exhibits 
the radial profile of the mean density $\bar{\rho}$ by solid lines.
The dot-dashed line indicates a critical density, $\rho$(sg,z), above 
which self-gravity of the disk in the vertical (z) direction must
be considered.
Lower two dashed lines mean the inertia of the heat, $\rho_{\rm rad}$,
for $\mdot =$100 and 1000.
The third panel depicts the mean temperature $\bar{T}$.
In the bottom, 
the effective temperature profiles $T_{\rm eff} (r)$ 
before the relativistic correction (solid lines)
are compared with those with the correction (dotted lines; eq.~\ref{eq:rc})
are shown.
Dashed line also shows $T_{\rm eff} (r)$, but it is the case
with no advection (i.e., $\Qvis = \Qrad$).
Due to the relativistic correction, 
the maximum $T_{\rm eff}$ of the slim disk is 
less than that of the $\Qvis = \Qrad$ case.
Dot--long-dashed lines in the third and forth panel present
the color temperature derived with eq.~\ref{eq_final_f_nu} for
the $\mdot = 1000$ case.
The spectral hardening factor 
can be estimated from the ratio of the color temperature to 
the effective temperature.
\label{fig:hdtt_m6_5}}
%
\noindent
Due to the relativistic correction, 
the maximum $T_{\rm eff}$ of the slim disk becomes
less than that of the $\Qvis = \Qrad$ case.
Dot--long-dashed lines in the third and forth panels present
the color temperature derived with eq.~\ref{eq_final_f_nu} for
the $\mdot = 1000$ case.

Since we shall use the spectra with $\alpha$ of 0.01 ($\Mbh = 10^{6.5} \Msun$)
when testing our model with 
the observed spectrum, we here briefly discuss the $\alpha$-dependency
of the flow.
The $\alpha$-dependencies of all the physical quantities 
(figures~\ref{fig:tau_m6_5}--\ref{fig:hdtt_m6_5})
are almost the same as the standard disk:
i.e., $\tau_{\rm es} \, (\propto \Sigma) \propto \alpha^{-1}$, 
$v_{\rm r} \propto \alpha$,
$t_{\rm acc}$ and $t_{\rm diff}$ are proportional to $\alpha^{-1}$, 
$H/r \propto \alpha^0$, $\bar{\rho} \propto \alpha^{-1}$, 
$\bar{T} \propto \alpha^{-0.25}$, etc.
The exception arises at the innermost region, $r \leq 10 \rs$.
Namely, $\tau_{\rm es} \, (\propto \Sigma)$, 
$v_{\rm r}$ and the two timescales
at $r \leq 1.5 \rs$ are unchanged.
The transition between the two different $\alpha$-dependencies occurs
at $1.5 \leq r \leq 10 \rs$.
The effective Compton-y parameters for the three values of $\alpha$ 
(with $\mdot = 1000$)
are drawn in the inset of fig~\ref{fig:y_m6_5}, showing
the decrease of $y_*$ at $r \lsim 4 \rs$ with decreasing $\alpha$.
Figure \ref{fig:hdtt_adep} depicts the aspect ratio (top), 
$\bar{\rho}$ (second), $\bar{T}$ (third), and $T_{\rm eff}$ (bottom)
for different $\alpha$ (0.1, 0.01, 10$^{-3}$)
with $\Mbh = 10^{6.5} \Msun$ and $\mdot = 1000$.
The behavior of $\bar{T}$ of the inner region 
is the opposite to that of the outer part.
Lower $\alpha$ makes the self-gravity of the flow more
important at outer radii.
The sonic point approaches the center with decreasing $\alpha$,
as shown in Abramowicz \ea (1988).
With smaller $\alpha$, $\bar{T} (r)$ tends to have a hump,
so that the pressure gradient [$d \Pi / dr \propto d(H \bar{T}^4)/dr$] 
pushes the gas towards the central BH more strongly.
The $\alpha$-dependency of
$T_{\rm eff} (r)$ is shown in Watarai \& Mineshige (2001).

\vspace{0.5cm}
\figurenum{6}
\centerline{\includegraphics[angle=0,width=8.5cm]{f6.epsi}}
\figcaption{\footnotesize
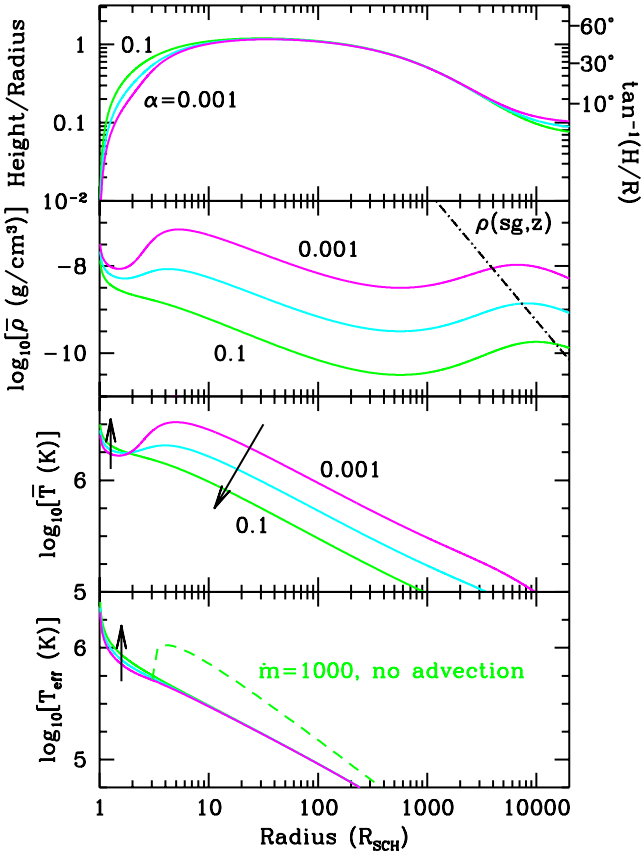
Aspect ratio (top), 
$\bar{\rho}$ (second), $\bar{T}$ (third), and $T_{\rm eff}$ (bottom)
 for different $\alpha$ (0.1, 0.01 and 10$^{-3}$)
with $\Mbh = 10^{6.5} \Msun$ and $\mdot = 1000$.
The $\alpha$-dependency of the physical quantities 
are almost the same as the standard disk,
except for the innermost region, $r \lsim 10 \, \rs$.
The behavior of $\bar{T}$ of the inner region 
with varying $\alpha$ is the opposite to that of the outer part.
Arrows mean the evolution with increasing $\alpha$.
\label{fig:hdtt_adep}}
\vspace{0.5cm}

Computations for $\Mbh = 32 \Msun$ are also made in order to
1) assess the $\Mbh$ dependency,
2)  apply the present model to ULXs/microquasars,
and 3) compare our spectra with those in Shimura \& Takahara (1995)
in the sub-Eddington regime.
Although the $\mdot$-dependency of the quantities 
(figures~\ref{fig:tau_m6_5}--\ref{fig:hdtt_m6_5}) in the slim disk 
is different from
the standard disk, the $\Mbh$-dependency is 
almost identical: e.g., 
$H/r \propto \Mbh^0$, 
$\bar{\rho}$\,(at the $p_{\rm rad}$-dominated region) $\propto \Mbh^{-1}$,
$\bar{T}$\,($p_{\rm rad}$ region) and $T_{\rm eff}$ 
are proportional to $\Mbh^{-1/4}$,
t$_{\rm acc}$ and t$_{\rm diff}$\,($p_{\rm rad}$ region) 
are proportional to $\Mbh$, etc.
The radius (in the unit of $\rs$) at which $p_{\rm rad} = p_{\rm gas}$ 
is proportional to $\Mbh^{2/21}$.

Different values of the ratio
$(\bar{\kappa_{\rm abs, 0}} / \bar{\kappa_{\rm ff, 0}})$
are used in different $\Mbh$ cases (\S 2.1),
thus absorption-dominated region is located farther out
than for $\Mbh = 10^{6.5} \Msun$.
Due to the change of the absorption coefficient, 
it gets easier for the flow to have $\tau_{\rm eff}$ less than unity:
$\tau_{\rm eff} \leq 1$ at $r \leq 10 \rs$ for $\mdot = 100$,
and at $r \leq 3 \rs$ for $\mdot = 1000$.
Comparing with 10$^{6.5} \Msun$ black holes, 
$y_*$ is enlarged by a factor of $\sim 100$
at all radii and all the $\mdot$.
The changes of $\bar{\rho}$ and $\bar{T}$ due to the decrease 
of $\Mbh$ by 5 orders result only in 0.6 order increase of $y_*$ 
(eq.~\ref{eq_y_mdot_dep}).
The rest ($\sim 1.5$ orders) arises from the decrease of 
$(\bar{\kappa_{\rm abs, 0}} / \bar{\kappa_{\rm ff, 0}})$.
Since $\rho ({\rm sg, z}) \propto \Mbh^{-2}$,
self-gravity of the disk around 32 $\Msun$ black holes is
not important at all.
Instead, it gets more significant with larger $\Mbh$.
Again, $\rho_{\rm rad}$ is always negligible.

\subsection{Emergent Spectra}

Figure \ref{fig:compare_b} shows the each effect of spectral calculation 
on the emergent spectrum for the case of $\mdot = 1000$ and $\alpha = 0.1$.
First, the dotted line represents the case with no advection: 
all the dissipated energy goes into radiation as in the
standard accretion disks.
For this curve, we put the inner radius at $3 \rs$.
The long-dashed line exhibits the spectrum 
with advection based on transonic accretion flow 
(i.e., the same as Mineshige \ea 2000, except for face-on view here).
We note that the local spectrum is still assumed as blackbody radiation here.
Next, we take into account the relativistic correction at 
the innermost region of the flow (dot-dashed line).
The flow extends inside the last stable orbit ($3 \rs$; 
e.g., Abramowicz \ea 1988; Watarai \ea 2000; Mineshige \ea 2000).
However, the emission from those regions, where gravity is 
extraordinarily strong
and azimuthal velocity is relativistic,
is drastically suppressed by the gravitational redshift
and transverse Doppler redshift.
As the result, the color temperature of the inner region
(where the highest energy photons are emitted)
decreases by a factor of $\sim 5.7$.
We should stress here that the bolometric luminosity of 
the flow $L$ is not very large in spite of the super-Eddington
accretion rate; $L \sim 2.6 \Ledd$ for the relativistic
corrected spectrum while it is 5.1 $\Ledd$ for 
the non-corrected spectrum.

Next, we add the effect of electron scattering on 
opacity (i.e., the modified blackbody spectrum; short-dashed line).
Finally, Comptonization is included in the computations (solid line).
Radiation from the inner region is boosted towards higher energy
by a factor of $\sim 3.4$ 
(the ratio of the color temperature to the effective temperature, 
called as the ``spectral hardening factor''; 
see. e.g., Shimura \& Takahara 1995) 
in comparison with the dot-dashed line.
This factor can be estimated if one compares 
the color temperature (dot--long-dashed line) and the effective 
temperature (solid line) in 
the bottom panel of fig.~\ref{fig:hdtt_m6_5}.
With the effects of electron scattering (opacity and Comptonization),
we get more gradual slopes in soft X-ray compared with the original
spectra (dotted line and long-dashed line).
The effect of Comptonization is included in the slim disks 
for the first time here (see also Shimura \& Manmoto 2003 
and Wang \& Netzer 2003).

\vspace{0.5cm}
\figurenum{7}
\centerline{\includegraphics[angle=0,width=8.5cm]{f7.epsi}}
\figcaption{\footnotesize
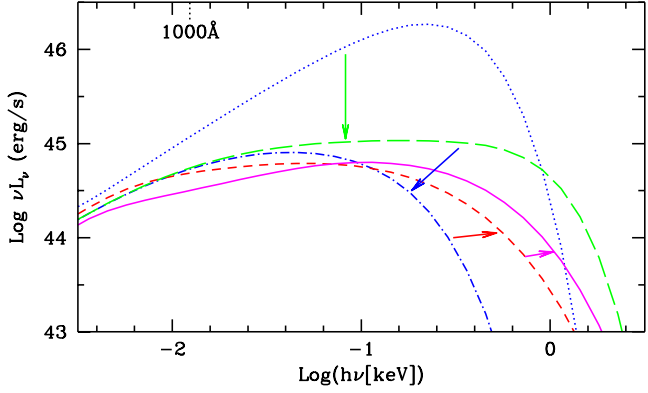
Each effect 
on the emergent spectrum for $\mdot = 1000$.
The dotted line represents the case with no advection 
(as in the
standard accretion disks) with an assumed 
inner radius at $3 \rs$.
Then, the long-dashed line exhibits the spectrum 
with advection based on the transonic accretion flow 
(i.e., the same as Mineshige \ea 2000, except for face-on view here),
and the bolometric luminosity of the flow $L$ is about 5.1 $\Ledd$.
The local spectrum is still assumed as blackbody radiation here.
Next, we take into account the relativistic correction at 
the innermost region of the flow (dot-dashed line).
As the result, the color temperature of the emission 
from the innermost region decreases by a factor of $\sim 5.7$,
and $L$ becomes to be 2.6 $\Ledd$.
More, we add the effect of electron scattering on 
opacity (i.e., modified blackbody spectrum; short-dashed line).
Radiation from the inner region is boosted towards higher energy.
Finally, Comptonization is included in the computations (solid line).
Comparing the solid and dot-dashed lines, 
the spectral hardening factor 
(the ratio of the color temperature to the effective temperature) 
at the innermost region is about 3.4.
\label{fig:compare_b}}
\vspace{0.5cm}

For the spectrum with all the effects ($\mdot = 1000$ and $\alpha = 0.1$), 
the contribution from
each radial component is shown in figure \ref{fig:contri}; 
$r \geq 10^3$, $r = 10^3$--100, 
100--10, 10--5, 5--3 and $r \leq 3 \rs$, 
respectively (thin solid lines).
For comparison, we also draw the spectra without 
the effects of electron scattering (but with the relativistic correction) 
by thin dot-dashed lines.
The spectral hardening factor of 
each spectrum is 1.3, 1.7, 2.3, 2.9, 3.4 and 4.0 for each radial region,
respectively.
This means that the inner emission suffers more spectral shift by
electron scattering.

\vspace{0.5cm}
\figurenum{8}
\centerline{\includegraphics[angle=0,width=8.5cm]{f8.epsi}}
\figcaption{\footnotesize
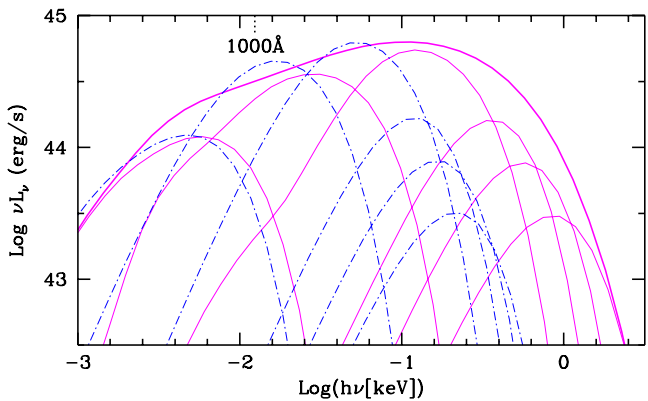
Contributions from different radii; $r \geq 10^3$, $r=10^3$--100, 
100--10, 10--5, 5--3 and $r \leq 3 \rs$, from left to right respectively,
for $\Mbh = 10^{6.5} \Msun$, $\mdot = 1000$ and $\alpha = 0.1$.
Thick solid line shows the spectrum from the entire radii 
with all the spectral effects as same as in fig.~\ref{fig:compare_b}, 
while 6 thin-solid lines mean the spectra from the different
radial regions above.
For comparison, we also draw the spectra without 
the effects of electron scattering (but with the relativistic correction) 
by thin dot-dashed lines.
The ``spectral hardening factor'' of 
each spectrum is 1.3, 1.7, 2.3, 2.9, 3.4, 4.0 for each radial region,
respectively.
\label{fig:contri}}
\vspace{0.5cm}

Figure \ref{fig:compare_adep} shows the $\alpha$-dependency of 
the emergent spectra with 
$\alpha$ = 0.1, 0.01 and 10$^{-3}$.
The arrows indicate the change of spectra with increasing $\alpha$.
The meanings of two kinds of lines (solid and dot-dashed) are 
the same as fig.~\ref{fig:compare_b}.
Without the effects of electron scattering, the spectra
are almost identical.
Spectral boosting by electron scattering is quite $\alpha$ sensitive.
Since $\tau_{\rm es} / \tau_{\rm abs} \propto \alpha^{1/8}$,
an increase of $\alpha$ results in an increase of the surface temperature
$T$ (eq.~\ref{eq_final_f_nu}), so we get more spectral boosting.
Previous models with the effect of electron scattering in opacity 
(Szuszkiewicz \ea 1996; Wang \ea 1999) 
employ a low viscosity parameter $\alpha =0.001$, in order to
guarantee that the flow is effectively optically thick.
Such a low $\alpha$ means that the spectral distortion by 
electron scattering is very small (fig.~\ref{fig:compare_adep}).

\vspace{0.5cm}
\figurenum{9}
\centerline{\includegraphics[angle=0,width=8.5cm]{f9.epsi}}
\figcaption{\footnotesize
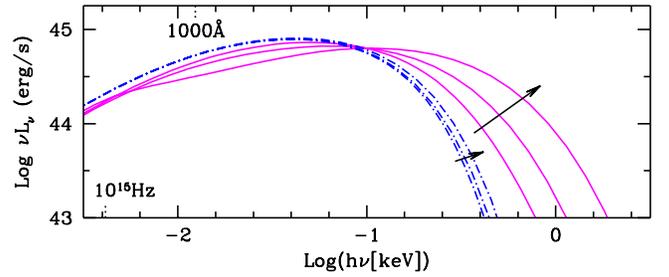
Spectra with three different values of $\alpha$:
$\alpha$ = 0.1, 0.01 and 10$^{-3}$.
The arrows indicate the change of spectra with increasing $\alpha$.
The meanings of two kinds of lines (solid and dot-dashed) are 
the same as in fig.~\ref{fig:compare_b}.
Without the effects of electron scattering, the spectra
are almost identical.
Spectral boosting by electron scattering is quite $\alpha$ sensitive.
\label{fig:compare_adep}}
\vspace{0.5cm}

Figure \ref{fig:ulx_spectr} shows the emergent spectra with
$\Mbh = 32 \Msun$ and $\alpha = 0.1$.
Meanings of the lines are the same as in fig.~\ref{fig:compare_b}.
Those spectra can correspond to those of some ULXs/microquasars.
The spectral hardening factors are 
around 1,3, 1.9 and 6.5 for $\mdot = 1$, 10 and 1000, respectively.
For $\mdot = 100$, it is quite difficult to measure it, since
the spectrum is highly distorted (see \S 4.2 for the reason).
The $\mdot$-dependency presented here is qualitatively the same as
the 10$^{6.5} \Msun$ results.
Some more details are shown in \S 4.2.
Due to the reduction of 
$(\bar{\kappa_{\rm abs, 0}} / \bar{\kappa_{\rm ff, 0}})$ in
the 10$^{1.5} \Msun$ calculations (\S 2.1),
the effects of electron scattering upon spectral boosting
are enhanced relative to 10$^{6.5} \Msun$ results (fig.~\ref{fig:compare_b}
and fig.~\ref{fig:pg1448_spectr}).
Shimura \& Takahara (1995) derived the spectral hardening factors to be
$\sim 1.7$ for $\mdot = 1$ and $\sim 1.9$ for $\mdot = 10$,
using the radiative transfer computations including Comptonization.
It is often assumed that the vertical distribution of the heating rate is
proportional to the volume gas density 
(Ross, Fabian \& Mineshige 1992; Shimura \& Takahara 1993,1995).
Our results are consistent with their values,
and thus the method used here for spectral calculations seems to 
work quite well.
Strong disk Comptonization, which is turned on at very high state in
a couple of microquasars, is suggested by Kubota (2001) 
and Kubota, Makishima \& Ebisawa (2001b).
We note that the inferred bolometric luminosity assuming the 
standard $T_{\rm eff} (r)$ profile (e.g., Mitsuda \ea 1984) will
be underestimating the true disk luminosity (perhaps by a factor of 1.5--2), 
since super-Eddington accretion flows have flatter
(in $\nu L_\nu$ plot) spectra than the standard accretion disk 
(fig.~\ref{fig:ulx_spectr}).

\vspace{0.5cm}
\figurenum{10}
\centerline{\includegraphics[angle=0,width=8.5cm]{f10.epsi}}
\figcaption{\footnotesize
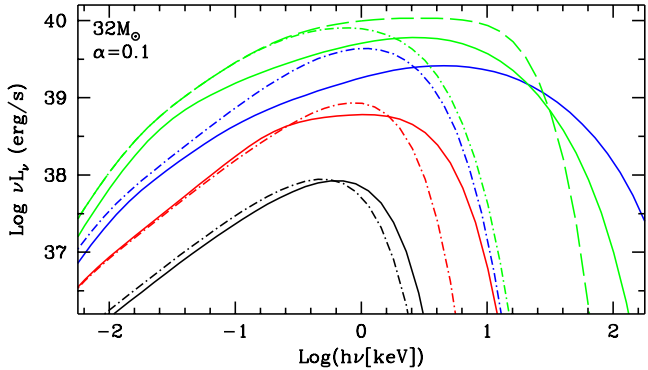
Emergent spectra from the face-on slim disks (solid lines) 
with $\Mbh = 32 \Msun$ and $\alpha = 0.1$.
Accretion rates $\mdot$ used for the computations are 
1, 10, 100 and 1000 from bottom to top, respectively.
The meanings of lines are 
the same as in fig.~\ref{fig:compare_b}.
Roughly, the spectral hardening factor is 1.3, 1.9 and 6.5 for 
$\mdot =$ 1, 10 and 1000, respectively.
Shimura \& Takahara (1995) reported that the spectral hardening 
factor $\simeq 1.7$
for $\mdot = 1$, and $\simeq 1.9$ for $\mdot = 10$ ($\Mbh = 1.3-10 \Msun$),
with a weak $\Mbh$-dependency ($\sim \Mbh^{0.1}$).
\label{fig:ulx_spectr}}
\vspace{0.5cm}

\section{OBSERVATIONAL TESTS OF THE MODEL}

Here, we will test the models with the data of AGNs, 
for which detailed data at both optical/UV and X-ray bands 
are available,
and some sorts of $\Mbh$ estimations have been developed.

\subsection{The Object Ideal for The Test}

Objects ideal for the test should have the highest $\dot{m}$, 
since we expect the largest difference in spectral shape 
from the spectrum of the standard accretion disk.
Also, the high $\dot{m}$ indicates 
a geometrically thick accretion flow (figure \ref{fig:hdtt_m6_5}),
then the self-obscuration may happen (Fukue 2000).
If we observe the innermost region of those highest $\mdot$ objects
(detections with soft X-ray will indicate it),
it implies that we are looking at the 
flow almost face-on.
Thus, we can neglect the longitudinal Doppler effect (Doppler boosting), 
which is an annoying effect in the model fitting.
[We simply estimate the luminosity as $4 \pi d_{\rm L}^2 \times$flux,
neglecting the inclination of the disk/flow.
The inferred luminosity 
will overestimate the true luminosity somewhat 
in the case of a concave flow-geometry
due to the self-irradiation (Fukue 2000; Misra \& Sriram 2003).]
Another reason to favor a large $\mdot$ is that 
the effective optical depth of the flow with $\mdot$ of 100
can be smaller than 1, so the radiative cooling rate
(eq.\ \ref{eq:qrad}) would not be accurate in that case.
We should select even higher $\mdot$ objects.
Another requirement for ideal test objects is that 
they should also harbor as small $\Mbh$ as possible.
Small enough $\Mbh$ allows
the accretion disk to become hot enough for
the emission from the innermost part of the flow 
to be easily observable in the soft X-ray band.

In terms of the temporal analysis, low $\Mbh$ selection is also
desirable in order to recognize the temporal behavior 
characteristic of the slim disk.
The accretion timescale at $r \lesssim$ 10 and $\lesssim$ 30 
R$_{\rm Sch}$ will have 
the order of a typical exposure time of X-ray satellite ($\sim$ 10 k~sec)
for $\dot{m}$ = 100 and 1000, respectively 
(Figure~\ref{fig:t_scales_m6_5}).
Thus, the slim disk is expected to change its dynamics
within one observation.
To select the candidates 
for high $\dot{m}$ and low $M_{\rm BH}$ objects, 
we plot {\it B}-band luminosity 
of each active nuclei $\nu L_\nu$($B$) as a function
of $M_{\rm BH}$ as follows.

There are several ways to estimate $M_{BH}$ of AGNs. 
(1) the reverberation mapping technique provides physically 
meaningful $M_{BH}$ estimation 
[$M_{BH}$(rev); e.g., Kaspi et al. 2000], though
the method is quite time consuming (an order of a year).
Now, $M_{BH}$(rev) estimations for AGNs whose FWHM(H$\beta$) less 
than 2000 km/s have been obtained for 9 objects 
(
Kaspi et al. 2000).
(2) An easier way is the one assuming some photo-ionization models 
for H$\beta$ lines [$M_{BH}$(ph); Wandel, Peterson \& Malkan 1999].
The two estimations agree with each other quite well.
The method utilizes the regression line between 
the size of broad line regions $R_{\rm BLR}$
and $\nu L_\nu$(5100 \AA),
determined in the reverberation mapping (Kaspi \ea 2000):
$R_{\rm BLR} = 32.9 \, [\nu L_\nu(5100$\AA
$)/(10^{44} {\rm erg/s})]^{0.7}$\,lt-days.
(3) Nelson (2000) and Ferrarese \ea (2001) 
found that the [O III] line width 
or the stellar velocity dispersion, $\sigma$, is 
correlated with $M_{BH}$(rev) in AGNs, 
and that they overlap the bulge mass--to--BH mass 
correlation in non-active galaxies (Gebhardt et al.\ 2000; 
Ferrarese \& Merritt 2000).
Thus, the [O III] width 
also provides an estimate of $\Mbh$, $M_{BH}$([O III]).
We use the regression reported as the equation 2 in Wang \& Lu (2001),
that was derived from $M_{BH}$(rev) or $M_{BH}$(ph) v.s.\ [O III] width
of the narrow-line and broad-line AGNs:
$M_{BH}$([O III]) $= 10^{7.78} \, [FWHM({\rm [O III]}) / 2.35 
/ (200\, {\rm km/s})]^{3.32} \, \Msun$.
We confirmed that an alternative regression (eq.~3) in Wang \& Lu (2001) 
does not change the result significantly.

There is some scatter around the regression lines,
both in the $R_{\rm BLR}$ v.s.\ $\nu L_\nu$(5100\AA) diagram
and in the $M_{BH}$ v.s.\ [O III] width plot.
To minimize uncertainty, 
we estimate $M_{\rm BH}$ from 
the geometric mean of the three estimations, namely
the square root of the product of $M_{BH}$(rev) and $M_{BH}$([O III]),
or that of $M_{BH}$(ph) and $M_{BH}$([O III]).
In general, those evaluations agree with each other quite well:
$\langle (\log_{10} \Mbh ({\rm rev \ or \ ph}) - 
\log_{10} \Mbh ({\rm [O III]}))^2 \rangle \,  = 0.54$.
The SDSS Early Data Release spectra also show a similar
scatter: $\sigma \simeq 0.67$ in $\log_{10}\Mbh ({\rm ph})$ (Boroson 2003).
In some objects without two available estimations, we use either of the three
\noindent\footnote{$M_{BH}$([O III]) for IC~3599 and 2E~1346+2646 and 
$M_{BH}$(rev) for PG~0052+251, PG~1226+023, PG~1426+015, PG~1617+175 
and PG~1700+518.}.

For some NLS1s, the [O III] line can be deconvolved into two components
(V\'{e}ron-Cetty, V\'{e}ron, \& Goncalves 2001, and references therein),
namely the broad, blueshifted one and the narrower one with the same
redshift as other lines.
The latter seems to arise 
from clouds showing normal galactic rotation.
For $M_{BH}$([O III]), the latter [O III] width is used 
as Wang \& Lu (2001) did,
if the [O III] line is deconvolved into the two components.
However, for I~Zw~I, it is known that neither of the two components
appear at the galaxy redshift 
(Phillips 1976; V\'{e}ron-Cetty et al.~2001): both components seem to show
outflows.
Therefore, this object is omitted in our plot.

In general, NLS1s have a narrower [O III] line width ($\lsim$ 300 km/s) 
than BLS1s.
To take account of a possible overestimation of the [O III] width 
due to the moderate spectral resolution ($\sim 200$km/s),
Wang \& Lu (2001) divide the [O III] width listed in 
V\'{e}ron-Cetty \ea (2001) by a factor of 1.3.
We follow the same procedure.
Spectroscopic observations 
with higher spectral resolution 
and, as for I~Zw~I, a careful analysis on the two [O III] components
will be required in the future to obtain a regression with smaller scatter.

\begin{figure*}[tb]
\figurenum{11}
\centerline{\includegraphics[angle=0,width=12cm]{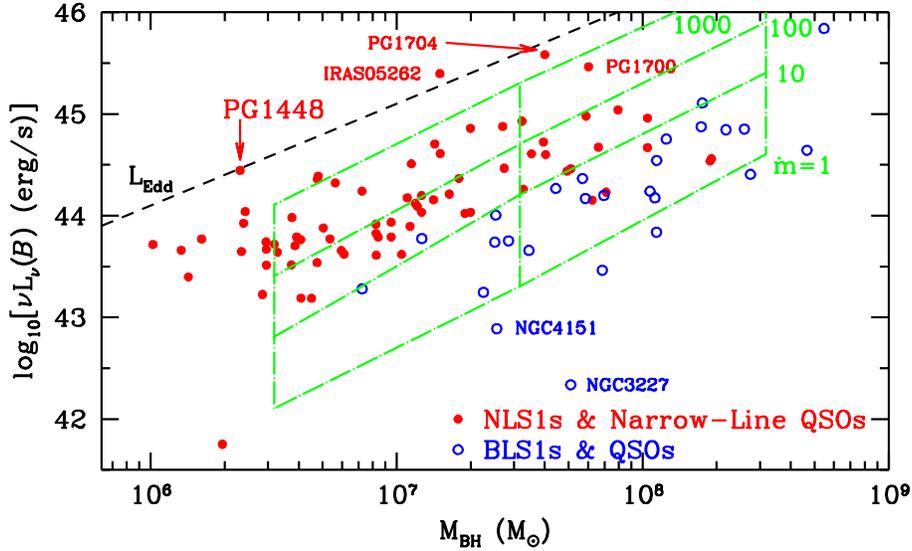}}
\figcaption{\footnotesize
B-band luminosity of active nuclei as a function of $M_{\rm BH}$.
Data are derived 
from Wang \& Lu (2001), Nelson (2000) and Kaspi et al.~(2000). 
NLS1s and narrow-line QSOs are denoted by filled circles, while BLS1s
and QSOs are plotted by open circles. 
Narrow-line objects show higher $\dot{m}$ than broad-line objects 
on average. 
Among the small $M_{\rm BH}$, higher $\dot{m}$ NLS1s, 
PG~1448+273 is the candidate with highest $\dot{m}$.
Then, this object can be desirable to test the overall 
slim disk model.
Dashed lines are loci computed by the model for different $\mdot$
with $\alpha = 0.1$.
Optical flux has a very weak $\alpha$-dependency (fig.~\ref{fig:compare_adep}).
Absolute $B$-band magnitude $M_B$ of -23 corresponds to $\nu L_\nu (B)$ 
of 10$^{44.8}$~erg/s.
\label{fig:lb_mbh_a}}
\end{figure*}

%
\begin{table*}[b]
\footnotesize
\caption{Additional NLS1s and a BLS1}
\begin{center}
\begin{tabular}{llrlrrllll}
\hline
\hline
Objects Name & 
z & 
H$\beta^a$ &
$m_V$ &
$A_V^b$ &
[O III]$^a$ & 
$L_B^c$ &
$R_{\rm BLR}$ &
$\Mbh^d$ &
$\Mbh^d$ \\
  &
  & 
FWHM &
(ref.) &
  &
FWHM & 
  &
(lt-day) &
(ph) &
([O III]) \\
  &
  &
(km/s) &
  &
  &
(km/s) &
  &
  &
  &
    \\
\hline
PHL 1092 \dotfill     & 0.396 & 1790 & 17.0\,(VV01)& 0.13 &       & 45.0 & 168.2 & 7.90 &     \\
RX J0439.7$-$4540\dotfill& 0.224 & 1010 & 16.6\,(G99) & 0.03 & 1020  & 44.7 & 92.2  & 7.14 & 8.90\\
NAB 0205+024\dotfill & 0.155 & 1050 & 15.4\,(K91) & 0.10 &       & 44.9 & 125.0 & 7.30 &     \\
\ \ (Mrk 586)  & & & & & & & & & \\
PKS 0558$-$504 \dotfill& 0.137 & 1250 & 15.0\,(R86) & 0.15 &       & 44.9 & 140.6 & 7.51 &     \\
1H 0707$-$495 \dotfill& 0.041 & 1050 & 15.7\,(R86) & 0.32&1516$^e$& 43.7 & 18.5  & 6.47 &     \\
IRAS 13224$-$3809\dotfill& 0.067 &650$^f$&15.2\,(Y99) & 0.23 &  810  & 44.3 & 47.8  & 6.47 & 8.56\\
IRAS 13349+2438\dotfill& 0.108 & 2200 & 14.3\,(M95) & 0.04 &       & 45.0 & 147.4 & 8.02 &     \\
IRAS 20181$-$2244\dotfill& 0.185 &  370 & 16.8\,(ES94)& 0.24 &  537  & 44.5 & 71.0  & 6.15 & 7.97\\
\hline
RX J0134.2$-$4258\dotfill& 0.238 &  900 & 16.0\,(G99) & 0.06 &  670  & 45.0 & 150.6 & 7.25 & 8.29\\
WPVS 007 \dotfill     & 0.029 & 1620 & 14.8\,(G99) & 0.04 &  320  & 43.6 & 17.0  & 6.81 & 7.23\\
RX~J2217.9-5941\dotfill& 0.160 & 1850 & 16.2\,(G99) & 0.08 & 1075  & 44.6 & 77.0  & 7.59 & 8.97\\
RX~J2349.3-3125\dotfill& 0.135 & 5210 & 16.6\,(G99) & 0.05 &  475  & 44.2 & 46.0  & 8.26 & 7.80\\
%
%
%
\hline
\end{tabular}
\vskip 2pt
\parbox{6in} 
{\small\baselineskip 9pt
\footnotesize
\indent
$^{\rm a}$ Line widths of H$\beta$ and [O III] are taken 
from Leighly (1999b) for the objects above the horizontal line,
and from Grupe \ea (1999), with $\sim$ 5\AA $\sim 300$ km/s resolution, 
for the lower objects. \\
$^{\rm b}$ Galactic extinction $A_V$ is derived from NED. \\
$^{\rm c}$ $\log_{10}(\nu L_{\nu}(B) [{\rm erg/s}])$ \\
$^{\rm d}$ $\log_{10}(\Mbh / M_\odot)$ \\
$^{\rm e}$ This [O~III] width is uncertain due to low [O~III] flux
and contamination from Fe~II blends (Leighly 2001, private communication).
Then, we estimate $\Mbh$ from $\Mbh$(ph) alone. \\
$^{\rm f}$ This H$\beta$ width may be underestimated, since there
seems some contamination of narrow H$\beta$ line arises from starburst 
of the host galaxy (Leighly 2001, private communication). \\
 {\sc REFERENCES. ---} 
G99: Grupe et al.\markcite{45} 1999; 
K91: Korista\markcite{65} 1991; 
R86: Remillard et al.\markcite{115} 1986; 
VV01: V\'{e}ron-Cetty \& V\'{e}ron 2001;
Y99: Young et al.1999;
M95: Mason et al.1995;
ES94: Elizalde \& Steiner1994
}
\end{center}
\vglue-0.9cm 
\end{table*}
\normalsize
%

In addition to the objects and data in Wang \& Lu (2001), Nelson (2000) 
and Kaspi \ea (2000), the following, well-known
NLS1s that are not tabulated in the literatures are added; 
namely, several NLS1s with the {\it ASCA} observations (see Leighly 1999b).
Some NLS1s show giant amplitude (a factor of $\sim$ 10--100) 
X-ray variability.
We also add such objects into our sample (see \S 6.1 for more details).
Table 1 lists the added objects; 8 objects from Leighly (1999b) and
4 from Grupe, Thomas \& Beuermann (2001).
In total, we have 100 objects; 74 narrow and 26 broad objects.
We note that the current sample is far from a complete sample
by any means.

The Galactic extinction is corrected using a standard
extinction curve (Savage \& Mathis 1979).
For all the conversion between optical bands and for the $K$-correction, 
it is assumed that $f_\nu \propto \nu^{-0.5}$ (e.g., Kaspi \ea 2000).
Accuracy of the optical photometric data must be considered carefully:
e.g., the nuclear fraction of the measured luminosity 
is $0.64-0.97$ for $\nu L_\nu$($B$)$\geq 10^{44.2}$erg/s sources
(Surace, Sanders \& Evans  2001).
Less luminous sources would contain less
nuclear fraction.
The optical variability is of smaller amplitude than X-ray variability.

Figure \ref{fig:lb_mbh_a} exhibits 
the $\Mbh$ v.s.\ $\nu L_\nu$($B$-band) plot.
This plot will be useful to estimate the order of $\dot{m}$ 
(roughly $\propto L_{\nu}(B)/M_{\rm BH}$) in a sample of objects.
The closer to the line of $L_{\rm Edd}$ (dashed line), 
the higher $\dot{m}$ such objects will have.
NLS1s/narrow-line QSOs seem to have
 higher $\dot{m}$ than BLS1s/QSOs.
If we do not apply the [O III]--width correction 
by a factor of 1/1.3 for the Wang \& Lu (2001) sample, most NLS1s 
(54 out of 74) shift to the right (larger $\Mbh$) by 0.2 dex.
Still there is the systematic difference of $\dot{m}$
between NLS1s/narrow-line QSOs and BLS1s/QSOs.
The objects' names are indicated for several outliers.
PG~1700 and PG~1704 are also pointed out as the highest $\mdot$ objects 
by Collin \ea (2002) who estimated $\mdot$ of 34 AGNs from
the optical luminosity and $\Mbh$(rev).

The best, high $\dot{m}$, low $M_{\rm BH}$ target is 
PG~1448+273 (z = 0.065).
The inferred BH masses are 
$M_{\rm BH}$(ph) = 8.4$\times$10$^6$ $M_\odot$ and 
$M_{\rm BH}$([O~III])  = 6.3$\times$10$^5$ $M_\odot$.
We thus estimate $\Mbh$ as $10^{6.4}$ $M_\odot$.
It has $\nu L_{\nu}(B) = 10^{44.4}$ erg/s. 
The FWHM width of H$_{\beta}$ and [O~III] are 820--1050 km/s and 155 km/s, 
respectively (Stirpe 1990; 
V\'{e}ron-Cetty et al.~2001).
The optical spectrum 
shows no evidence for 
contamination from the host galaxy
into the optical luminosity, and it is not expected in 
{\it B}-band (Stirpe 1990).
These widths are still narrower than typical NLS1s, indicating 
that this object will be one of the extreme NLS1s.
We expect the object will show the hottest accretion disk 
among the sample,
which will enable us to investigate the nature of accretion flow 
unprecedentedly well with soft X-ray observations ({\it XMM-Newton}).
No X-ray observations have been performed for this object
except for ROSAT All Sky Survey (RASS); 
the 0.1--2~keV photon index $\Gamma$ is 
3.2$\pm$0.3~(1$\sigma$),
 and $\nu F_{\nu}$(2~keV) is 1.4$\pm$0.5
10$^{-12}$ erg/s/cm$^2$ (Walter \& Fink 1993).
There are no special remarks on this object in Grupe \ea 
(2001) where some objects showing strange variability during RASS
observations are reported.

\subsection{Comparison of the Models with the Available Data}

Now, we compare our model spectra (\S 3.2) with the currently available,
observed data of PG~1448+273 at optical--Soft X-ray bands 
(Figure \ref{fig:pg1448_spectr}).
The standard disk model ($\mdot \lsim 10$) 
does not fit the broadband spectrum of 
PG~1448+273 (dot and a tie-bow error), and thus 
a slim disk model is required. 
The Soft X-ray photon index of PG~1448+273 ($\Gamma \sim 3.2$) 
is larger than normal BLS1s (see e.g., Boller \ea 1996), 
but is not the largest one
among the NLS1s ($\Gamma \sim 2.5 - 4.5$).

Figure \ref{fig:pg1448_spectr} 
exhibits a comparison of spectra between the optical ({\it B}-band) 
-- soft X-ray (by RASS) observations and the models.
Here, $\Mbh$ is taken to be $10^{6.5} \Msun$, 
and $\alpha = 0.1$ (top) or $\alpha = 0.01$ (bottom).
The corresponding $\mdot$ value for each line is shown in
the figure.
The spectral hardening factors for $\alpha = 0.1$ cases are
$\sim$ 1.2, 1.3 and 3.4 for $\mdot = 1$, 10 and 1000, respectively,
while $\alpha = 0.01$ results indicate the factor to be $\sim$ 2.8 and 2.3
for $\mdot = 100$ and 1000, respectively.
The $\mdot = 1000$ and $\alpha = 0.1$ model 
with all the effects (solid line) 
has a problem: the soft X-ray spectrum is too 
boosted towards higher energy to explain the observed value.
We found that 
the PG 1448$+$273 data is well fitted by a model with an extremely large
accretion rate; $\mdot = 1000$ and $\alpha = 0.01$.
As far as the soft X-ray spectrum is concerned, both of
the $\mdot =$ 100 and 1000 models with $\alpha = 0.01$ can fit the data.
Thus, 
it is very important to look at both of optical/UV and soft X-ray bands.
We note that most of the dissipated energy 
is advected and swallowed into the BH as trapped photons.
This is the first example where optical--soft X-ray spectral distribution
of NLS1s are explained by models. 
The dashed lines are the curves based on the previous computation 
of the slim disk
(Mineshige et al.\ 2000) with $\dot{m}=1000$,
which turned out to have difficulty in explaining the 
soft X-ray luminosity/spectral slope.
In the bottom panel, spectra with an additional 
hard X-ray power-law are also shown, for presentation purpose. 
The photon index of 2.15 (Brandt \ea 1997; Leighly 1999b), 
with the 3000\AA--2~keV spectral index of 1.6 
(Brandt, Laor \& Willis 2000) are used.

Figure \ref{fig:pg1448_spectr} 
also exhibits the $\Mdot$ dependence of the model spectra 
with a fixed $\Mbh$.
We recall that $\tau_{\rm eff} \leq 1$ at the inner region, with 
$\mdot = 100$ and $\alpha = 0.1$.
Because of the exponential term in eq.~\ref{eq_final_f_nu} 
(the effect of the finite $\tau_{\rm eff}$),
the surface temperature $T$ drastically increases
with the parameter set.
The spectral hardening factor, the degree of spectral boosting,
 relative to the model without electron scattering is
indeed enhanced and becomes more strongly radius-dependent 
(see Shimura \& Takahara 1995, \S=3.4 for $r$-dependency of 
the factor in sub-Eddington AGN disks)
in the $\mdot = 100$ case than the $\mdot = 1000$ result.
Up to $\mdot \leq 10$, the peak frequency of the disk emission 
($T_{\rm in}$)
varies with $\Mdot$ in a similar fashion to 
that of the standard accretion disk: $\Tcol \propto 
 \Mbh^{-1/2} \Mdot^{1/4} 
 \ \propto \ \Mbh^{-1/4} \mdot^{1/4}$.
Above that accretion rate, there is a jump in $T_{\rm in}$,
due to the $\mdot$-sensitive spectral boosting via electron scattering.
Such a jump does not happen without electron scattering effects 
(dot-dashed lines).

\vspace{0.5cm}
\figurenum{12}
\centerline{\includegraphics[angle=0,width=8.5cm]{f12.epsi}}
\figcaption{\footnotesize
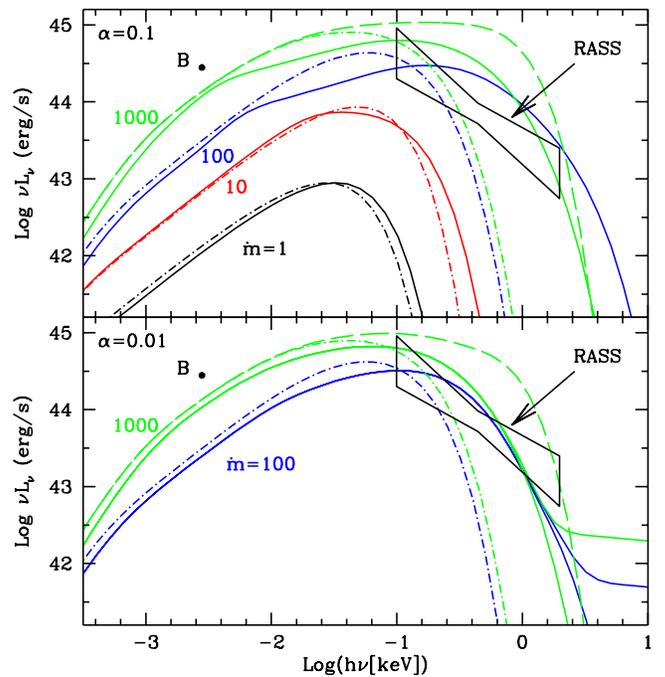
Spectral models of the face-on slim disk (solid lines), 
with $\Mbh = 10^{6.5} \Msun$,
for the observed data of PG~1448+273 (a dot
and a tie-bow; errors indicate 1~$\sigma$ of $\Gamma$ and flux).
Accretion rate used for the computations are labeled.
Bolometric luminosities $L$ are 
0.021, 0.21, 1.2 and 2.6 of $L_{\rm Edd}$ for 
$\mdot =$ 1, 10, 100 and 1000, respectively. 
The meanings of lines are 
the same as in fig.~\ref{fig:compare_b}.
The dot-dashed and long-dashed lines are computed 
without the effects of electron scattering: the 
relativistic correction is taken into account
in the former, while not in the latter.
Top: The cases with $\alpha = 0.1$.
Roughly, the spectral hardening factor is 1.2, 1.3 and 3.4 for 
$\mdot =$ 1, 10 and 1000.
(For $\mdot = 100$ it is difficult to measure it, since the 
spectrum is highly distorted. See the text for that reason.)
To reproduce the optical flux, $\mdot \geq 1000$ is required.
The $\mdot = 1000$ model with all the effects (solid line) 
has a problem: the soft X-ray spectrum is too 
boosted towards higher energy to explain the observed value.
Bottom: The same as above, except for $\alpha = 0.01$ here.
Models for $\mdot =$ 100, 1000 are shown,
with the resultant spectral hardening factor of 
$\sim 2.8$ and 2.3, respectively.
For presentation purpose, we add a hard X-ray power-law 
with a photon index of 2.15 for each spectrum, with the 
3000\AA--2~keV spectral index of 1.6.
It is found that the $\mdot = 1000$ case reproduce the observed 
broad-band spectrum quite well.
\label{fig:pg1448_spectr}}
\vspace{0.5cm}

With $\mdot \geq 100$, $T_{\rm in}$ saturates, as found 
in previous works (Wang \ea 1999; Watarai \ea 2000; Mineshige \ea 2000).

As shown above, we have established that the slim disk model is promising 
for PG~1448+273 for B-band and RASS data
($\dot{m}=1000$ producing 2.6$\times L_{\rm Edd}$).
This is the first success in reproducing optical--soft X-ray
emission simultaneously.
To verify and constrain our model further,
we need a much higher signal-to-noise ratio and 
a broader band pass obtainable with {\it XMM-Newton}.
We will show the detailed comparison between the
model and the data, which will be obtained by 
our approved observation (PI; T.~Kawaguchi), in a future paper. 
Also, whether we are able to fit the spectra at different flux states
with the same $M_{\rm BH}$ provides a strong constraints on the 
accretion models.

\subsection{Predicted X-ray Features}

Below, we state several features predicted by the model.
These properties will be judged by future observations.

{\it Is the spectrum distorted due to electron scattering?}\\
The most distinguishable 
feature of the slim disk is the soft X-ray spectrum (soft excess
or soft X-ray hump) 
from the innermost region of the disk.
The soft excess 
may not be fitted well by the so called multi-color disk spectra, 
since the slim disk model 
predicts that the soft X-ray emission is significantly distorted due to 
the strong radius-dependent, spectral hardening factor
 at a certain range of $\mdot$ 
($\mdot = 100$ with $\alpha = 0.1$; fig.~\ref{fig:pg1448_spectr} top panel).
Such a power-law-like soft excess 
was recently reported in a couple of NLS1s, 
Ton~S~180 (Vaughan et al.~2002) and Mrk~478 (Marshall \ea 2003) 
with
$\Gamma \, \sim 3$ and $\sim 1.5$ for soft and hard X-ray, respectively.

{\it Intrinsic spectral features}\\
The temperature of the innermost region 
of NLS1s with $\mdot \gsim 1000$ 
seems to be around $10^6$ K.
Those objects tend to have weaker corona than coronae of low $\mdot$
objects.
Then, it can be possible to observe the intrinsic spectral features
(bound-free and bound-bound) of the accretion flow in the soft X-ray band.
Such features have commonly been attributed to ``warm absorbers''.
However, the strong Comptonization (fig.~\ref{fig:y_m6_5}) may
smear those features (see Shimura 2000).
Some discussions upon possible 
disk emission features in {\it XMM} data
have appeared recently (Branduardi-Raymont \ea 2001; Mason \ea 2003).

{\it Does the color temperature change with flux?}\\
The color temperature of the radiation 
from the innermost disk ($T_{\rm in}$)
may vary with flux significantly
(Figure~\ref{fig:pg1448_spectr}).
Such a behavior of the slim disk 
(significant change of $T_{\rm in}$ associated with less
change in $L$)
is expected 
to occur at $\mdot =$ 10--100, 
due to the $\dot{m}$-sensitive electron scattering. 
If we detect the significant $T_{\rm in}$ change, 
distinct from 
the simplest (without the effects of electron scattering) 
standard disk model ($T_{\rm in} \propto L^{1/4}$), 
it is an evidence for strong electron scattering, 
and hence (see fig.~\ref{fig:y_m6_5}) it is an indication 
of super-Eddington accretion in NLS1s.

It should be stressed that 
a similar $T_{\rm in}$-change has been observed in ULXs/microquasars 
(see Makishima \ea 2000):
e.g., Mizuno et al.~(2001) found that three ULXs exhibit 
$T_{\rm in} \propto L^{1/2}$, which is different from the relation 
in the standard disk, $T_{\rm in} \propto L^{1/4}$.

{\it Is Photon trapping important or does convection dominate?}\\
Another expectation from the slim disk model is that 
the soft excess might 
be reduced at higher $\dot{m}$ due to advection (photon trapping).
The accretion timescale will become shorter than the timescale
of photon diffusion (fig.~\ref{fig:t_scales_m6_5}).
We may therefore expect less soft X-ray excess--to--optical flux ratio
for extremely high $\dot{m}$ than for a moderately large $\dot{m}$ case. 
That could be verified from the combination of OM and EPIC observations, 
since 
optical/UV radiation suffers less from the photon trapping.
Absence of this effect will mean that the convective energy 
transport in the disk is more efficient than the radiative one.

\section{GROWTH TIMESCALE OF BHS}

The elapsed time since the 
gas-fueling started
will have the order of $\Mbh/\Mdot \ (= 0.5 \, \mdot^{-1}$\,Gy) 
unless a sudden change of accretion rate occurs.
Thus, NLS1s and narrow-line QSOs could be the key objects 
for figuring out what turned
on the efficient gas-accretion towards the central BHs.
As described in the previous section,
we found that some of them, 
including PG 1448$+$273, have an extremely high
accretion rate: $\mdot \approx 1000$. 
This implies that the object is really young: its age 
inferred from $\Mbh / \Mdot$ is 
about $10^6$ years.
Due to the rapid growth of $\Mbh$, such a high $\mdot$ stage
will be a quite short phase.

It may be important to note that 
some ULXs exhibit extended optical nebulae (Pakull \& Mirioni 2002).
Assuming that a nebula is a remnant of a supernova-like event,
they estimate the age of one ULX as $\sim$ 1~Myr based on
the size of the nebula 
($\sim$ 400 pc in diameter), expansion velocity and H$\beta$ 
luminosity.

\section{DISCUSSION}

\subsection{Highly Variable AGNs}

Some NLS1s show giant amplitude (a factor of $\sim$ 10--100) 
X-ray variability;
short-term ($\lsim$ a day) variabilities in 
IRAS 13224-3809 (Boller et al. 1997; see also Otani et al. 1996),
PHL 1092 (Forster \& Halpern 1996; Brandt et al. 1999),
RX~J2217.9-5941 (Grupe et al.\ 2001),
and Mrk 766 (Leighly et al.1996; Grupe \ea 2001), 
as well as long-term ($\sim$ years) variabilities in 
NGC 4051 (Guainazzi \ea 1998; Uttley \ea 1999),
RE J1237+264 (= IC 3599; Brandt \ea 1995; Grupe \ea 1995a; 
 Komossa \& Bade 1999),
1H 0707-495 (Leighly et al.\ 2002) and
WPVS 007 (Grupe \ea 1995b), etc.
To investigate the physics and nature behind these phenomena,
we have added such objects into figure~\ref{fig:lb_mbh_a};
3 NLS1s, WPVS 007 (= 1RXS J003915.6-511701; Grupe \ea 1995b),
RX~J0134.2-4258 (Grupe et al.\ 2000),
RX~J2217.9-5941 
and one BLS1 RX~J2349.3-3125 (Grupe et al.\ 2001).
RX~J1304.2+0205, HS1702+32 (Grupe et al.\ 2001) are not included since
there are no optical data available.

When $\dot{m}$ goes up, 
a shorter accretion timescale is associated with 
a higher radial velocity 
(Figure~\ref{fig:vv_m6_5} and \ref{fig:t_scales_m6_5}).
This suggests that such specific NLS1s (and one BLS1) 
might have the highest $\dot{m}$ among all NLS1s.
Some of the highly variable and transient AGNs (e.g., NGC~4051)
may be relevant to obscuration (e.g., Brandt \& Gallagher 2000)

\begin{figure*}[tb]
\figurenum{13}
\centerline{\includegraphics[angle=0,width=12cm]{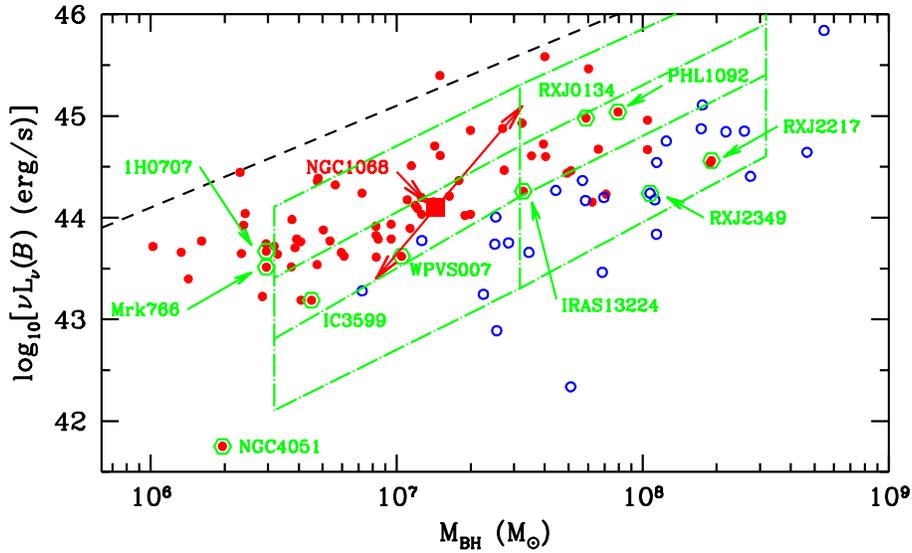}}
\figcaption{\footnotesize
Resultant distribution of highly variable/transient AGNs 
(hexagonal dots).
They do not belong to the highest $\mdot$ group.
Instead, it seems that such objects are located at 
the intermediate region where the two classes merge.
A large filled square denotes NGC~1068 (\S 6.3).
\label{fig:lb_mbh_b}}
\end{figure*}

The resultant distribution of those AGNs are shown in Figure 
\ref{fig:lb_mbh_b} together with other objects.
It turns out that they do not belong to the highest $\mdot$ group.
This is true even if we use one specific $\Mbh$ estimation among
the three, rather than the geometric mean.
Instead, it seems that such highly-variable objects are located at 
the intermediate region where the two classes merge 
($3 \lsim \mdot \lsim 300$).
If this is true, 
the distribution may indicate that such high-amplitude variability 
is linked with a transience between 
a standard disk and a slim disk 
due to the thermal instability 
(Honma et al. 1991a,b; Leighly et al.~2002).
Further studies with a sample including more transient NLS1s 
are required to examine this hypothesis.

Janiuk, Czerny \& Siemiginowska (2002) discuss the thermal instability
in terms of transitions of microquasars.
Transitions seen in some ULXs (Kubota \ea 2001a) can 
also be relevant to the instability.

\subsection{Photon Trapping}

As shown in fig.~\ref{fig:t_scales_m6_5}, 
the diffusion timescale of photons becomes longer than
the accretion timescale (photon trapping).
We expect that the diffusion approximation (eq.~\ref{eq:qrad}) 
should provide a reasonable estimation for the cooling rate,
although Ohsuga \ea (2002) argue that it might not always be true 
for super-Eddington accretion rates.
A similar trapping phenomenon can be seen for the neutrino trapping 
in a newly born neutron star (proto-neutron star) associated
with a Type II (core-collapse) supernova.
The diffusion timescale of neutrinos from the dense
core (seconds) is longer than the dynamical timescale
of collapse or bounce shock around the neutrino-sphere
(1--10 ms; see Burrows 1990 for a review).
Numerous studies (e.g., Janka \& Hillebrandt 1989; 
Mezzacappa \ea 1998, etc), adopting
the (flux-limited) diffusion approximation 
and Monte-Carlo simulations to treat the neutrino leakage, 
have been made for calculating 
the inner structure and 
light curve of the neutrino flux from a (convective) proto-neutron 
star.
The long history of this field would help to consider
the cooling rate in super-Eddington accretion flows.

The inner structure of a magnetized accretion flow will be highly inhomogeneous
(e.g., Machida \ea 2000; Kawaguchi \ea 2000).
Photons inside the flow tend to escape by passing through 
a low-density region (see Begelman 2001).
Therefore, the degree of photon trapping, and thus that of advection, 
might be reduced when we take this into account.
However, the photon diffusion timescale (dotted lines in 
fig.~\ref{fig:t_scales_m6_5}) is much longer than the dynamical
timescale ($\Omega_{\rm K}^{-1}$; dot-dashed line).
Since the inner inhomogeneity will change within the timescale of
$\Omega_{\rm K}^{-1}$, photons will still have difficulty in
escaping from the flow.
A lower $\alpha$ enhances the deviation of the two timescales.

\subsection{NGC 1068; a Type 2 NLS1?}

Here, we briefly discuss the accretion rate of 
a well-known Seyfert 2 galaxy, NGC~1068.
The relatively narrow $\Hb$ line and significant FeII multiplets in 
the polarized emission from NGC~1068 (Miller, Goodrich, Mathews 1991) 
lead people to suspect it
as a type 2 counterpart of NLS1s.


For NGC~1068, the mean [OIII] width ($\sim$1150km/s)
can not be used for $\Mbh$ determination, since NLR clouds seem to suffer
from the interaction with the jet (Dietrich \& Wagner 1998).
The lowest FWHM ($\sim$ 200km/s) 
[OIII] clouds seem to be consistent with rotation
around the nucleus,
and the FWHM width of the line complex from those clouds 
is about 210km/s (Dietrich \& Wagner 1998).
Consequently, the implied $M_{\rm BH}$([O~III]) is 
4.2 10$^{6} \Msun$. 
Also, its polarized H$\beta$ line width is $\sim$ 3000km/s 
(Miller \ea 1991), and its intrinsic
optical luminosity [$\nu L_{\nu}(B)$] was estimated as 
$\sim 10^{44.1} (f_{\rm refl}/0.01)^{-1}$ erg/s, 
assuming the fraction of nuclear flux reflected into
our line of sight $f_{\rm refl}$ to be 0.01 (Pier et al. 1994).
Therefore, $M_{\rm BH}$(ph) = 5.1$\times$10$^7 (f_{\rm refl}/0.01)^{-0.7}$ 
$M_\odot$.
The geometric mean is 10$^{7.2} (f_{\rm refl}/0.01)^{-0.35}$ $M_\odot$.
We note that $f_{\rm refl}$ is thought to be between 0.05 
(Bland-Hawthorn, Sokolowski, Cecil 1991) and 0.001 
(Bland-Hawthorn \& Voit 1993).
The inferred $\Mbh$ is consistent with the $\Mbh$ estimated from
water masers in the central pc, 
(1-2)$\times$10$^7 \Msun$ (Greenhill \& Gwinn 1997).

The filled square in Figure \ref{fig:lb_mbh_b} indicates
the location of NGC~1068,
showing that it indeed has a high $\mdot$ as NLS1s.
The error-bar shown here arises from the uncertainty of $f_{\rm refl}$ alone.

\subsection{Comparison with The QSO Composite Spectrum}

NLS1s are the dominant source of diffuse 
EUV photons in the local universe (Edelson et al.~1999).
To examine the contribution of NLS1s in the distant, young universe,
we need to understand the shape of the 
 EUV spectrum as functions of $M_{BH}$
and $\dot{M}$.

In order to recognize the difference in spectral energy distributions
between NLS1s and QSOs, we plot the QSO composite spectrum 
by dotted lines 
(Zheng et al.\ 1997; Laor et al.\ 1997; Telfer \ea 2002) 
upon our model spectra 
($\Mbh = 10^{6.5} \Msun$, $\mdot = 100$ and $1000$ with $\alpha = 0.01$)
in figure~\ref{fig:pg1448_spectr_qso}. 
Following Laor \ea (1997), 
the vertical normalization of the composite spectrum is chosen
to match with a representative NIR luminosity of nearby
PG quasars, but re-adopting a Hubble constant of 75
km/s/Mpc (namely we divide by 2.25 from the spectrum in figure~6 
of Laor \ea 1997).
Although the $\mdot \sim 1000$ period will be a quite short phase
and thus rare,
the phase with $\mdot \sim 100$ will be common in NLS1s and narrow-line QSOs 
(fig.~\ref{fig:lb_mbh_a}).
As shown in figure~\ref{fig:pg1448_spectr_qso}, 
the two types of objects (NLS1s and QSOs)
show quite different distributions.
The ratio of the EUV luminosity to optical ones or 
EUV to X-ray ratio are really different in 
the high $\mdot$ systems and QSOs.
The EUV luminosity is comparable in the $\mdot = 1000$ case and 
the QSO spectrum, while the EUV spectral slopes are not similar.
The slope in EUV band plays an important role in indicating 
to what extent EUV photons penetrate into primordial gas clouds 
against self-shielding (e.g., Tajiri \& Umemura 1998).
Similar comparison can be seen when one plots the QSO composite 
spectrum onto the broadband spectrum of another NLS1, Mrk 478, in
Marshall \ea (1996).

\vspace{0.5cm}
\figurenum{14}
\centerline{\includegraphics[angle=0,width=8.5cm]{f14.epsi}}
\figcaption{\footnotesize
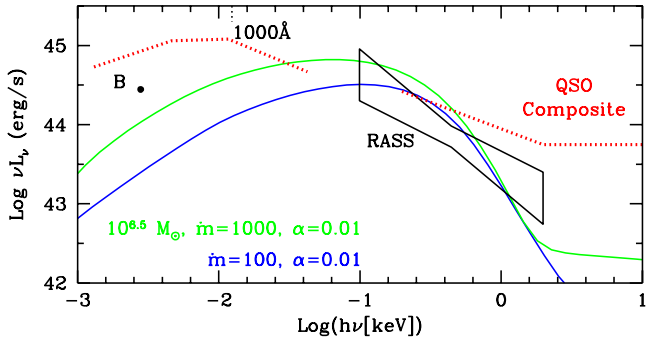
Comparison between the spectral models for NLS1s [solid lines; 
$\mdot =$100 (lower) and 1000 (upper) 
with $\Mbh = 10^{6.5} \Msun$ and $\alpha = 0.01$] and the QSO 
composite spectrum 
(dotted lines; Zheng et al.\ 1997; Laor et al.\ 1997; Telfer \ea 2002).
Observational data for PG~1448+273 are also shown for a presentation
purpose.
The two types of objects 
show quite different distributions.
Extreme-UV luminosity is comparable in a NLS1 case ($\mdot = 1000$) and 
the QSO case, while optical and X-ray 
ones are so different.
\label{fig:pg1448_spectr_qso}}
\vspace{0.5cm}

\subsection{Other Perspectives}

As prescribed in \S 5, NLS1s can be in the younger phase of central
BH formation/evolution.
Mathur (2000) discuss the scenario in the context of gas content
and metal richness of some NLS1s.
If the ``young AGN'' hypothesis is the valid, NLS1s provide us a good
sample for studying the evolution and formation of massive BHs 
residing in the center of galaxies.
Krongold, Dultzin-Hacyan \& Marziani (2001) compared the environments
of NLS1s with those of BLS1s based on the Digitized Sky Survey,
finding no statistical difference between the two.
We will need much better spatial resolution to observe 
 disturbances on host galaxies that may feed BHs and ignite 
AGN activity, such as a ripple structure, 
tidal tail, nucleus bar and so on.
Our forthcoming observation of nearby NLS1s with The University of 
Hawaii 2.2-meter telescope (PI; K.~Aoki) will
assess the validity of this idea.

With recent intensive studies, 
it has been established that mass of the galactic spheroid component 
(galactic bulge or elliptical galaxy itself) is strongly correlated 
with $\Mbh$ in in-active galaxies and AGNs
(Gebhardt et al.\ 2000; 
Ferrarese \& Merritt 2000; Nelson 2000).
The tight correlation indicates that the growth of spheroid mass 
(in other words, formation of spheroids) and growth of $\Mbh$
(formation of BHs) are triggered by the same 
origin, or it implies that either controls the other growth/formation.
The origin of the link is still unclear.
Two, opposite, extremal ideas can be that 
i) Jet/Outflow due to gas accretion
onto the central BH may govern the formation of spheroid (Silk \&
Rees 1998), i.e. BHs are made before the bulge growth;
and 
ii) radiation drag by massive stars during the
bulge formation can fuel the gas towards
the central BH (Umemura 2001), resulting in bulges formed 
prior to BH growth.
Deep, optical/Near-IR, off-nuclear spectroscopic 
observations of spheroids (bulges) associated with the youngest
 AGNs will reveal the origin of the close correlation.
The time-lag between the epochs of the bulge-- and BH--growth
could be recognized, 
depending on the spectral features of the bulges we shall obtain.
It will promote 
insights into why the two massive, but with enormously 
different physical sizes, systems co-evolve.

\section{Summary}

We have examined the effects of 
electron scattering (opacity and Comptonization) 
and the relativistic correction (gravitational redshift and
transverse Doppler effect)
on the emergent spectra from super-Eddington accretion flows 
(the so called slim disks) onto non-rotating BHs
for $10^{1.5}$ and $10^{6.5} \Msun$ $\Mbh$. 

The effective optical depth of the flow
can be less than unity for $\mdot \, 
[\equiv \Mdot / (\Ledd / c^2)]\, = 100-1000$ with 
$10^{1.5} \Msun$ and for $\mdot = 100$ with $10^{6.5} \Msun$,
if $\alpha \geq 0.1$.
An improvement in the knowledge of the cooling rate 
will be required in those parameter sets.
Although $\mdot$-dependency and radial profile of each physical
quantity in the super-Eddington flow are different from the standard disk,
the $\Mbh$-dependency of the two accretion modes are almost identical.
The $\alpha$-dependency is also similar, except for the innermost
region ($r \lsim 10 \rs$).
narrow-line QSOs will be desirable laboratories for studying
the self-gravity of the flow via optical/Near-IR observations.

With $\mdot \geq 100$, the spectral hardening factor
 via electron scattering is $\sim 2.3 - 6.5$.
The color temperature of the inner most radiation ($\Tin$)
is not proportional to $L^{0.25}$, as would be expected from the 
simplest standard accretion disk, due to the $\mdot$-sensitive 
spectral hardening factor.
In the face-on view, there is no jump of $\Tin$ if we do not take the 
$\mdot$-sensitive electron scattering into account.

The improved spectral model has been applied to 
the optical--soft X-ray emission from NLS1s, 
for which $\Mbh$ are now estimated by 
a couple of ways,
to evaluate their accretion rates $\mdot$. 
We have picked up one NLS1, PG~1448+273 with an inferred $\Mbh$ of 
$10^{6.4} \Msun$, among the highest $\mdot$ candidates.
The broadband spectral distribution was
successfully reproduced for the first time by the model 
with $\mdot = 1000$ and $\alpha = 0.01$. 

This extremely high $\mdot$ 
implies that this object, as well as some other highest $\mdot$
systems, is really young: 
its inferred age ($\Mbh / \Mdot$) is less than $10^6$ years.

We have also briefly discussed the accretion rate $\mdot$ 
of transient and highly variable NLS1s, finding that
those are located at $3 \lsim \mdot \lsim 300$.
Such a moderately high accretion rate can indicate that
this variability is relevant to the thermal instability.
A possible type 2 counterpart of NLS1s, NGC~1068, seems to 
have $\mdot$ similar to NLS1s.

A more accurate estimate for absorption opacity,
for which we assume 30 times free-free opacity in this study for simplicity, 
is required for better spectral calculations.
It is because the absorption opacity determines the location
of the last thermalization surface, and hence the 
amount of redistributed photons by Comptonization.
Also, viewing-angle dependent spectral models are needed.

\acknowledgements
This paper could never have been produced in this way without the 
use of the numerical code for solving the radial structure of 
super- and sub-Eddington accretion flows/disks
that has been 
constructed and developed by Ryoji Matsumoto,
Fumio Honma and Mitsuru Takeuchi.
The author wish to thank 
Karen Leighly, 
Chiho Matsumoto, 
Ken-ya Watarai,
Kentaro Aoki,
Toshiya Shimura, 
Agata R\'{o}\.{z}a\'{n}ska,
Jerzy Madej,
Bozena Czerny,
Ed Baron, 
Jean-Marc Hur\'{e},
Suzy Collin, 
Chris Done and
Omer Blaes
for helpful discussions.
He also thanks
Brandon Carter for patient and careful reading of the draft, 
Shin Mineshige for continuous encouragement, and
an anonymous referee for careful reading and helpful comments.
The main computations were performed at the Department of 
Physics and Astronomy of the University of Oklahoma,
when the author was there as a visiting graduate student.
He appreciates all the member of the Department, especially
C.\ Matsumoto and K.\ Leighly, for their hospitality.
This research has made use of the NASA/IPAC Extragalactic Database (NED) 
which is operated by the Jet Propulsion Laboratory,
California Institute of Technology, under contract with the National 
Aeronautics and Space Administration. 
This work was supported in part by the Research Fellowship of the Japan
Society for the Promotion of Science (JSPS) for Young Scientists (4616), 
and by the JSPS Postdoctoral Fellowships for Research Abroad (464).






\begin{thebibliography}{}
\bibitem[]{1916}
 Abramowicz M.A., Czerny B., Lasota J.P. \& Szuszkiewicz E., 1988. ApJ,
 332, 646
\bibitem[]{1919}
 Begelman M.C. \& Meier D.L., 1982, ApJ, 253, 873
\bibitem[]{1921}
 Begelman M.C., 2001, ApJ, 551, 897
\bibitem[]{1923}
 Beloborodov A.M. 1998, MNRAS 297, 739
\bibitem[]{1925}
 Beloborodov A.M., Abramowicz M.A. \& Novikov I.D. 1997, ApJ, 491, 267
\bibitem[]{1927}
 Bland-Hawthorn J., Sokolowski J. \& Cecil G., 1991, ApJ, 375, 78
\bibitem[]{1929}
 Bland-Hawthorn J. \& Voit G.M., 1993, Rev. Mex. Astron. Astrofis., 27, 73
\bibitem[]{1931}
 Boller Th., Brandt W.N., Fink H.H. 1996, A\&A 305, 53
\bibitem[]{1933}
 Boller Th., Brandt W.N., Fabian A.C., Fink H.H. 1997, MNRAS 289, 393
\bibitem[]{}
 Boroson, T. A. 2003, ApJ, 585, 647
\bibitem[]{1935}
 Brandt, W. N., Pounds, K. A., \& Fink, H 1995, 273, L47
\bibitem[]{1937}
 Brandt W.N. \& Gallagher S.C., 2000, New Astronomy Reviews, 44, 461
\bibitem[]{1939}
 Brandt W.N., Mathur S., Elvis M. 1997, MNRAS 285, L25
\bibitem[]{1941}
 Brandt W.N., Laor A. \& Wills B.J., 2000, ApJ, 528, 637
\bibitem[]{1943}
 Brandt W.N., Boller Th., Fabian A.C., Ruszkowski 1999, MNRAS 303, L53
\bibitem[]{}
 Branduardi-Raymont, G., \ea 2001, A\&A, 365, L140
\bibitem[]{1945}
 Burrows A.S., 1990, in Supernovae (Springer-Verlag) ed. A.G. Petschek, p.143
\bibitem[]{1947}
 Cheng L., Wei J. \& Zhao Y., 2002, Chinese J. Astron. Astrophys., 2, 207
\bibitem[]{1949}
 Colbert E. \& Ptak A., 2002, ApJS, 143, 25
\bibitem[]{1951}
 Collin, S. et al.\ 2002, A\&A, 388, 771
\bibitem[]{1953}
 Czerny, B. \& Elvis, M. 1987, ApJ, 321, 305
\bibitem[]{2150}
 Dietrich M. \& Wagner S.J., 1998, A\&A, 338, 405
\bibitem[]{1955}
 Edelson R. et al., 1999, MNRAS, 307, 91
\bibitem[]{1957}
 Elizalde F. \& Steiner J.E., 1994, MNRAS, 268, L47
\bibitem[]{1959}
 Ferrarese L.\ \ea 2001, ApJL, 555, 79
\bibitem[]{1961}
 Ferrarese L. \& Merritt D. 2000, ApJ, 539, L9
\bibitem[]{1963}
 Forster, K. \& Halpern, J.P., 1996, ApJ, 468, 565
\bibitem[]{1965}
 Fukue J. 2000, PASJ, 52, 829
\bibitem[]{1967}
 Gebhardt K. et al.\ 2000, ApJ, 539, L13
\bibitem[]{1969}
 Goldreich P. \& Lynden-Bell D., MNRAS, 1965, 130, 97
\bibitem[]{1971}
 Greenhill L.J. \& Gwinn C.R., 1997, Ap\&SS, 248, 261
\bibitem[]{1973}
 Grupe, D., Beuermann, K., Mannheim, K., Bade, N., Thomas, H.-C., 
 de Martino, D. \& Schwope, A. 1995a, A\&A, 299, L5
\bibitem[]{1976}
 Grupe, D., Beuerman, K., Mannheim, K., Thomas, H.-C., Fink, H. H., 
 de Martino, D. 1995b, A\&A, 300, L21
\bibitem[]{1979}
 Grupe G.D., Beuermann K., Thomas H.-C., Mannheim K. Fink H.H. 
	  1998, A\&A 330, 25
\bibitem[]{1982}
 Grupe G.D., Beuermann K., Mannheim K., Thomas H.-C. 1999, 
	  A\&A 350, 805
\bibitem[]{1985}
 Grupe, G.D., Thomas H.-C. \& Beuermann K., 2001, A\&A, 367, 470
\bibitem[]{1987}
 Grupe G.D., Leighly K.M., Thomas H.-C., \& Laurent-Muehleisen S.A.
 2000, A\&A, 356, 11
\bibitem[]{1990}
 Guainazzi, M., \ea 1998, MNRAS, 301, L1
\bibitem[]{1992}
 Halpern, J. P., \& Oke, J. B. 1987, ApJ 312, 91
\bibitem[]{1994}
 Hayashida, K. 2000, New Astron. Rev., 44, 419
\bibitem[]{1996}
 Honma F., Matsumoto R. \& Kato S., 1991, PASJ, 43, 147
\bibitem[]{1998}
 Honma F., Matsumoto R., Kato S. \& Abramowicz M.A., 1991, PASJ, 43, 261
\bibitem[]{2000}
 H\={o}shi R., 1977, Prog. Theor. Phys., 58, 1191
\bibitem[]{2002}
 Hur\'{e} J.-M., 1998, A\&A, 337, 625
\bibitem[]{2004}
 Janiuk A., Czerny B. \& Siemiginowska A., 2002, ApJ, 576, 908
\bibitem[]{2006}
 Janka H.-T. \& Hillebrandt W., 1989, A\&AS, 78, 375
\bibitem[]{2008}
 Kaspi S., Smith P.S., Netzer H., Maoz D., Jannuzi B.T. \& Giveon U. 
	  2000, ApJ, 533, 631
\bibitem[]{2011}
 Kato S., Fukue J., \& Mineshige S. 1998, Black Hole Accretion Disks, 
	  Kyoto Univ. Press
\bibitem[]{2014}
 Kawaguchi, T., Mineshige, S., Machida, M., Matsumoto, R., \& Shibata,
	  K. 2000, PASJ, 52, L1
\bibitem[]{}
 Kawaguchi, T., Pierens, A., \& Hur\'{e}, J.-M. 2003, A\&A, submitted
\bibitem[]{2017}
 Korista K.T., 1991, AJ, 102, 41
\bibitem[]{2019}
 Krongold Y., Dultzin-Hacyan D. \& Marziani P., 2001, AJ, 121, 702
\bibitem[]{2021}
 Kubota A., 2001, Ph.D. thesis, Univ. Tokyo
\bibitem[]{2023}
 Kubota A.\ et al.\ 2001a, ApJ, 547, L119
\bibitem[]{2025}
 Kubota A., Makishima K. \& Ebisawa K., 2001b, ApJ, 560, L147
\bibitem[]{2027}
 Komossa S. \& Bade N. 1999, A\&A, 343, 775
\bibitem[]{2029}
 Laor, A., \& Netzer, H. 1989, MNRAS, 238, 897
\bibitem[]{2031}
 Laor A., Fiore F., Elvis M., Wikes B. J. \& McDowell J. C. 1997, 
	  ApJ, 477, 93
\bibitem[]{2034}
 Leighly, K. M. et al.\ 1996, ApJ, 469, 147
\bibitem[]{2036}
 Leighly, K. M. 1999a, ApJS, 125, 297
\bibitem[]{2038}
 Leighly, K. M. 1999b, ApJS, 125, 317
\bibitem[]{2040}
 Leighly, K. M., Zdziarski, A. A., Kawaguchi, T., \& Matsumoto, C. 2002, 
 in X-ray Spectroscopy of AGN with Chandra and XMM-Newton, 
 ed. Th. Boller, S. Komossa, S. Kahn, \& H. Kunieda 
 (MPE Rep. 279; Garching: MPE), 259--262 (astro-ph/0205539)
\bibitem[]{2043}
 Liu B.F., Mineshige S., Meyer F., Meyer-Hofmeister E. \& Kawaguchi T., 
2002, ApJ, 575, 117
\bibitem[]{2046}
 Machida M., Hayashi M.R. \& Matsumoto R., 2000, ApJ, 532, L67
\bibitem[]{}
 Madej J., 1974, Acta.\ Astronomica, 24, 327
\bibitem[]{2048}
 Marshall H.L. et al., 1996, ApJ, 457, 169
\bibitem[]{}
 Marshall H.L., \ea, 2003, AJ, 125, 459
\bibitem[]{}
 Mason, K. O., \ea, 2003, ApJ, 582, 95
\bibitem[]{2050}
 Matsumoto R., Kato S., Fukue J., Okazaki A.T. 1984, PASJ 36, 71
\bibitem[]{2052}
 Matsumoto R., Kato S. \ Honma F., 1989, in Theory of Accretion Disks,
 eds. F. Meyer, W.J. Duschl, J. Frank, and E. Meyer-Hofmeister 
 (Kluwer Academic Publishers, Dordrecht), p.167
\bibitem[]{2056}
 Mathur S., 2000, MNRAS, 314, L17
\bibitem[]{2058}
 Makishima K., et al.\ 2000, ApJ, 535, 632
\bibitem[]{2060}
 Mason K.O. et al., 1995, MNRAS, 274, 1194
\bibitem[]{2062}
 Meyer, F., \& Meyer-Hofmeister, E., 1994, A\&A, 288, 175
\bibitem[]{2064}
 Mezzacappa A. et al.\ 1998, ApJ, 493, 848
\bibitem[]{}
 Miller J.M., Fabbiano G., Miller M.C. \& Fabian A.C., 2003, 
 ApJL, 585, 37
\bibitem[]{2066}
 Miller J.S., Goodrich R.W. \& Mathews W.G., 1991, ApJ, 378, 47
\bibitem[]{2068}
 Mineshige S., Kawaguchi T., Takeuchi M. \& Hayashida K. 2000, 
	  PASJ, 52, 499
\bibitem[]{2071}
 Misra R. \& Sriram K., 2003, ApJ, 584, 981
\bibitem[]{2073}
 Mitsuda K., et al., 1984, PASJ, 36, 741
\bibitem[]{2075}
 Mizuno T., Kubota A. \& Makishima K., 2001, ApJ, 554, 1282
\bibitem[]{2077}
 Narayan R., Yi I. 1995, ApJ 444, 231
\bibitem[]{2079}
 Narayan R., 1997, in Accretion Phenomena and Related Outflows, Proc.\ IAU 
 Colloq.\ 163, ASP Conf.\ series (eds. D.T. Wickramasinghe, 
G.V. Bicknell, L. Ferrario), p.75
\bibitem[]{2083}
 Nelson C.H. 2000, ApJL, 544, L91
\bibitem[]{2085}
 Ohsuga K., Mineshige S., Mori M. \& Umemura M., 2002, ApJ, 574, 315
\bibitem[]{2087}
 Otani C., Kii T., Miya K. 1996
 in R\"ontgenstrahlung from the Universe
 (MPE Report 263), ed H.U. Zimmermann, J.E. Tr\"umper, H. Yorke
  (MPE Press, Garching) p491
\bibitem[]{2092}
 Osterbrock D.E., \& Pogge R.W.\ 1985, ApJ, 297, 166
\bibitem[]{2094}
 Paczy\'{n}ski B., Wiita P.J. 1980, A\&A 88, 23
\bibitem[]{2096}
 Pakull, M. W., \& Mirioni, L. 2002,
 in New Visions of the X-ray Universe in the XMM-Newton and Chandra Era, 
 in press (astro-ph/0202488)
\bibitem[]{2098}
 Phillips M.M., 1976, ApJ, 208, 37
\bibitem[]{2100}
 Pier E.A. et al.\ 1994, ApJ, 428, 124
\bibitem[]{2102}
 Pogge R.W.\ 2000, New Astronomy Reviews, 44, 381
\bibitem[]{2104}
 Pounds K.A., Done C., Osborne J. 1995, MNRAS 277, L5
\bibitem[]{2106}
 Roberts T.P. \& Warwick R.S., 2000, MNRAS, 315, 98
\bibitem[]{2108}
 Ross, R. R., Fabian, A. C., \& Mineshige, S. 1992, MNRAS, 258, 189
\bibitem[]{2110}
 Rybicki, G. B., \& Lightman, A. P. 1979, Radiative Processes in
	  Astrophysics, John Willey \& Sons, New York
\bibitem[]{2113}
 Remillard R.A. et al., 1986, ApJ, 301, 742
\bibitem[]{2115}
 Savage B.D. \& Mathis J.S., 1979, ARA\&A, 17, 73
\bibitem[]{2117}
 Shakura N.I. \& Sunyaev R.A., 1973, A\&A, 24, 337
\bibitem[]{2119}
 Shimura, T., \& Takahara, F. 1993, ApJ, 419, 78
\bibitem[]{2121}
 Shimura, T., \& Takahara, F. 1995, ApJ, 445, 780
\bibitem[]{}
 Shimura, T., 2000, MNRAS, 315, 345
\bibitem[]{}
 Shimura, T., \& Manmoto, T. 2003, MNRAS, 338, 1013
\bibitem[]{2123}
 Silk J. \& Rees M., 1998, A\&A, 331, L1
\bibitem[]{2125}
 Stirpe G.M., 1990, A\&AS, 85, 1049
\bibitem[]{2127}
 Surace J.A., Sanders D.B. \& Evans A.S., 2001, AJ, 122, 2791
\bibitem[]{2129}
 Svensson R., 1984, MNRAS, 209, 175
\bibitem[]{2131}
 Szuszkiewicz, E., Malkan, M. A., \& Abramowicz, M. A. 1996, ApJ, 458, 474
\bibitem[]{2133}
 Tajiri Y., \& Umemura M., 1998, ApJ, 502, 59
\bibitem[]{2135}
 Takeuchi, M. 2000, Poster contribution presented at the Joint 
 MPE/AIP/ESO workshop on
 Observational and Theoretical Progress in the Study
 of Narrow-line Seyfert 1 Galaxies (astro-ph/0005162)
\bibitem[]{2137}
 Telfer R.C. et al.\ 2002, ApJ, 565, 773
\bibitem[]{2139}
 Umemura M., 2001, 2001, ApJ, 560, L29
\bibitem[]{2141}
 Uttley, P., McHardy, I.M., Papadakis, I. E., Guainazzi, M. \& Fruscione, A., 
1999, MNRAS, 307, L6
\bibitem[]{2144}
 Vaughan S. et al.\ 2002, MNRAS, 337, 247
\bibitem[]{2146}
 V\'{e}ron-Cetty M.P., V\'{e}ron P. \& Goncalves A.C., 2001, A\&A, 372, 730
\bibitem[]{2148}
 V\'{e}ron-Cetty M.P. \& V\'{e}ron P., 2001, A\&A, 374, 92
\bibitem[]{2152}
 Walter R. \& Fink H. H. 1993, A\&A, 274, 105
\bibitem[]{2154}
 Wandel A., Peterson B.M. \& Malkan M.A. 1999, ApJ, 526, 579
\bibitem[]{2156}
 Wandel A. \& Petrosian V. 1988, ApJ, 329, L11
\bibitem[]{2158}
 Wang T., Brinkmann W. \& Bergeron J. 1996, A\&A, 309, 81
\bibitem[]{2160}
 Wang J.-M., Szuszkiewicz E., Lu F.-J. \& Zhou Y.-Y., 1999, ApJ 522, 839
\bibitem[]{2162}
 Wang J.-M. \& Netzer H., 2003, A\&A, 398, 927
\bibitem[]{2164}
 Wang T. \& Lu Y., 2001, A\&A, 377, 52
\bibitem[]{2166}
 Watarai K., Fukue J., Takeuchi M. \& Mineshige S., 2000, PASJ, 52, 133
\bibitem[]{2168}
 Watarai K., Mizuno T. \& Mineshige S., 2001, ApJ, 549, L77
\bibitem[]{2170}
 Watarai K. \& Mineshige S., 2001, PASJ, 53, 915
\bibitem[]{2172}
 Young A.J. et al., 1999, MNRAS, 304, L46
\bibitem[]{2174}
 Zheng, W. Kriss, G.~A., Telfer, R.~C., Grimes, J.~P., \&
	  Davidsen, A.~F. 1997, ApJ, 475, 469
\end{thebibliography}
\end{document}